\documentclass[twocolumn,amssymb,nobibnotes,aps,prx,superscriptaddress]{revtex4-2}       

\usepackage{graphicx}
\usepackage{amsmath, amsthm, amssymb}
\theoremstyle{definition}

\usepackage{mathdots}
\usepackage{float}
\usepackage{pgfplots} 
\usepackage{tikz-3dplot}
\usepgfplotslibrary{patchplots}
\usepgfplotslibrary{groupplots}
\pgfplotsset{compat=1.6,ylabsh/.style={every axis y label/.style={at={(0,0.5)}, xshift=#1, rotate=90}}}
\usetikzlibrary{positioning}
\usetikzlibrary{shapes}
\usetikzlibrary{shapes.geometric}
\usetikzlibrary{decorations.text}
\usetikzlibrary{decorations.pathreplacing,shapes.misc}
\usetikzlibrary{matrix}
\usepackage{caption,subcaption}
\usepackage[ruled,lined]{algorithm2e}
\usepackage{algorithmic}
\renewcommand{\algorithmiccomment}[1]{\bgroup\hfill//~#1\egroup}
\usepackage{xcolor}
       \newenvironment{frcseries}{\fontfamily{frc}\selectfont}{}  

\begin{document}
\title{Active tissue mechanics drives a diversity of collective flows in uniaxially compressed 3D spheroids}

\author{Liyang Wang} \affiliation{School of Chemistry and Chemical Engineering, Shanghai Jiao Tong University, Shanghai 200240, China}

\author{Shabeeb Ameen} \affiliation{Physics Department, Syracuse University, Syracuse, NY 13244, USA}

\author{J. M. Schwarz}
\affiliation{Physics Department, Syracuse University, Syracuse, NY 13244, USA}
\affiliation{Indian Creek Farm, Ithaca, NY 14850, USA}

\author{Tao Zhang} \email{zhangtao.scholar@sjtu.edu.cn} \affiliation{School of Chemistry and Chemical Engineering, Shanghai Jiao Tong University, Shanghai 200240, China}

\date{\today}

\begin{abstract}
How compression affects the organization of collective cell motion remains unclear given the range of experimental outcomes, from enhanced or suppressed translational motion to the emergence or absence of global rotational motion.  Using a minimal three-dimensional active vertex model, we examine how the mechanical state of the tissue and its active dynamics in terms of strength and alignment shape the response of confined spheroids to uniaxial compression. In the absence of alignment, compression enhances transverse motion in solid-like spheroids throughout the tested activity range, whereas for the fluid-like response there is a change from enhanced transverse motion at smaller activity to decreased transverse motion at high activity. Changes in contact persistence via cell rearrangements help drive these responses. Specifically, local five-cell, passive calculations reveal how compression reshapes local rearrangement barriers: the response depends on geometry and loading direction, and a high-strain increase in the fluid-like mean coexists with many low-barrier paths. In the large-alignment regime, strong compression induces coherent rotation in solid-like spheroids. Fluid-like spheroids, on the other hand, show no comparable collective rotation. The rotating solids have a higher contact-persistence index than the fluids and exhibit tangential peripheral cell organization. Controlled perturbations reveal compression-induced softening of a differential-twist mode and slower relaxation with activity. These results identify an activity-dependent reversal of the motility response and compression-induced coherent rotation as two distinct outcomes of loading, linking them to contact persistence, local rearrangement energy costs, and active forces. Importantly, our results demonstrate that a minimal model, with parameters representing effective mechanical and active tissue properties, can capture qualitatively distinct collective responses to compression, motivating experiments to map cell types onto model parameters, enabling predictions for additional cell types.
\end{abstract}
\maketitle
\section{Introduction} \label{sec:intro}

Three-dimensional cell collectives exhibit a wide range of flows. In epithelial spheroids and organoids, cells can organize into coherent rotation: mammary acini rotate as they assemble \cite{tanner2012coherent,wang2013rotational}, pancreatic spheres rotate as active solids \cite{tan2024emergent}, and rotating mammary spheroids carry curvature-induced velocity waves \cite{brandstatter2023curvature}. Coherent rotation also underlies cell-chirality readouts \cite{chin2018epithelial} and egg-chamber elongation \cite{cetera2014epithelial}. Other collective behaviors include tumor-cell streaming into the surrounding extracellular matrix \cite{friedl2009collective,ilina2020cell} and disordered cellular flows \cite{krajnc2018fluidization,parker2025does,zhang2025}. Describing these behaviors requires understanding both how strongly cells move and how their motion is organized across the aggregate.

Mechanical loading can change both aspects of motion. Compression has been reported to suppress spheroid proliferation \cite{cheng2009micro} and enhance breast-cancer-cell migration and invasion \cite{tse2012mechanical,luo2022compression}. In collagen-embedded spheroids, Pandey \textit{et al.}\ found that compression increased MDA-MB-231 cell motility while reducing MCF10A motility \cite{pandey2024mechanical}. Under a different loading and imaging protocol, Suh \textit{et al.}\ observed compression-induced vortex motion and increased speed in MCF10A spheroids, whereas MDA-MB-231 spheroids slowed and showed no comparable vortex \cite{suh2025microfluidic}. Transient compression can also restore coherent rotation in developing malignant mammary structures \cite{ricca2018transient}. These observations establish that compression can alter both speed and collective organization, with outcomes that depend on experimental conditions. They motivate identifying the physical factors that determine the direction and spatial organization of the response.

Tissue mechanics offers a physical framework for understanding these contrasting responses. Compression--recovery measurements show that MCF10A spheroids recover more rapidly after unloading and exhibit less plastic deformation than the tumorigenic MCF7 and MDA-MB-231 spheroids examined in the same study \cite{pandey2026viscoelastic}. More generally, epithelia can exhibit jammed, solid-like or unjammed, fluid-like behavior, with cell shape serving as a key structural variable \cite{bi2015density,park2015unjamming,oswald2017jamming}. Activity can promote fluidization in an active two-dimensional Voronoi model \cite{bi2016motility}; unjamming accompanies collective motion in development and cancer \cite{mongera2018fluid,palamidessi2019unjamming}; and adhesion and matrix confinement influence jamming in breast-cancer invasion \cite{ilina2020cell}. Compressive stress can itself unjam cell sheets through adhesion strengthening \cite{cai2022compressive}. Yet persistent cell contacts can also support motion as an active solid \cite{tan2024emergent}. These findings highlight two roles of the contact network: constraining relative rearrangements and supporting coordinated motion. Two-dimensional active vertex models also combine active speed with local directional interactions to generate diverse migration patterns and collective flows \cite{lin2018dynamic,lin2021energetics}.

Our central question is therefore: How does active tissue mechanics determine whether compression enhances cell motion, suppresses it, or organizes it into coherent rotation? We address this question using a three-dimensional active vertex model of a confined spheroid \cite{zhang2025,Ameen2026, ho2026vimentin}. The target shape index $s_0$ controls the tissue's mechanical state \cite{Zhang2022}, which is also influenced by active speed and cell--cell alignment.  The same physical ingredients and dynamical rules underlie all full-spheroid simulations, allowing us to examine how distinct collective responses emerge for different parameters. We study two complementary regimes: an activity sweep with alignment switched off, and a separate high-alignment series focused on coherent rotation. Their distinct parameter sets and loading protocols are specified in Model and Methods.

The zero-alignment results reveal that activity can reverse the fluid-like response to compression even at a fixed target shape index. Fluid-like spheroids accelerate at low activity and slow at high activity, whereas solid-like spheroids accelerate throughout the tested range. Changes in contact persistence place these responses on a common empirical trend. Ordered and random five-cell calculations then examine the local mechanical cost of rearrangement. The random clusters reveal how compression reshapes the distribution of local barriers across geometries and loading directions, identifying heterogeneous mechanical constraints on rearrangement.

In the high-alignment regime, compression-induced coherent rotation coexists with greater neighbor retention than in the fluid-like spheroids. At $60\%$ strain, solid-like spheroids develop coherent rotation accompanied by tangential organization of peripheral cells, while fluid-like spheroids show no comparable collective rotation. Controlled perturbations demonstrate that compression softens a differential-twist mode and that activity slows its relaxation. These measurements motivate a heuristic deformation--polarization feedback mechanism. Together, the two regimes connect the magnitude and organization of motion to tissue mechanics and active dynamics. They yield two experimentally testable predictions: an activity-dependent reversal of the fluid-like compression response, and coherent rotation accompanied by a relatively persistent contact network.

\section{Model} \label{sec:model}

We represent a spheroid as a three-dimensional confluent cellular aggregate with vertex-model energy
\begin{equation}
\begin{aligned}
E_{\mathrm{VM}}={}&K_V\sum_j(V_j-V_0)^2+K_A\sum_j(A_j-A_0)^2\\
&+\Gamma\sum_f\delta_{f,B}A_f .
\end{aligned}
\label{eq:vertex_energy}
\end{equation}
Here $V_j$ and $A_j$ are the volume and surface area of cell $j$, $V_0$ and $A_0$ are their preferred values, and $K_V$ and $K_A$ are the corresponding stiffnesses. The last term assigns surface tension $\Gamma$ to boundary faces, selected by $\delta_{f,B}$. The target shape index $s_0=A_0/V_0^{2/3}$ controls the mechanical state: the zero-alignment comparison uses solid-like spheroids at $s_0=5.0$ and fluid-like spheroids at $s_0=5.6$. The high-alignment series uses solid-like spheroids at $s_0=4.8$ and fluid-like spheroids at $s_0=5.8$; the controlled twist assay uses the same solid-like target shape index. The two series have distinct parameter and loading protocols, specified in Methods. Here, $s_0$ specifies preferred area and volume rather than the actual cell shape index. The value $s_0=4.8$ lies slightly below the spherical lower bound $(36\pi)^{1/3}\simeq4.836$ for the actual ratio $A/V^{2/3}$. It therefore represents preferred area and volume values that cannot be satisfied simultaneously, rather than imposing a geometrically inaccessible cell shape.

Each cell carries a unit polarity $\boldsymbol\theta_j$. Here, polarity denotes the cell's active force direction, rather than an explicitly resolved morphological head--tail axis. As for what determines a cell's active force direction, or polarity, given that cells share interfaces, its evolution is driven by rotational noise with rotational diffusion constant $D_r$. Second, an alignment interaction with strength $\alpha$ acts directly on the active force directions of face-sharing neighboring cells, tending to coordinate their active driving. The relative importance of directional alignment and rotational diffusion is characterized by the dimensionless ratio \(\alpha/D_r\). Specifically, the high-alignment simulations use \(\alpha/D_r=3\times10^4\), placing them in a regime where alignment strongly dominates rotational diffusion. The evolution of the active force direction also includes separation and confinement steering. Separation turns the active force direction away from sufficiently close neighboring cells, while confinement turns it inward near the spheroid boundary as cells approach it. These interactions act on the directions of active driving rather than directly imposing forces on cell positions. Although these prescriptions are indeed minimal, they provide a framework for examining how different forms of cellular activity contribute to collective motion.

The overdamped velocity of a shared vertex $a$ combines mechanical forces, the active contributions of its incident cells, and noise due to active fluctuations:
\begin{equation}
\frac{d\mathbf r_a}{dt}
=\mu\big[-\nabla_a E_{\mathrm{VM}}+\mathbf F_a^{\mathrm{plate}}\big]
+v_0\sum_{j\ni a}\boldsymbol\theta_j+\boldsymbol\zeta_a(t).
\label{eq:vertex_dynamics}
\end{equation}
The coefficient $\mu$ denotes mobility, $\mathbf F_a^{\mathrm{plate}}$ is the force from the compression plates, and $v_0$ sets the active contribution per incident cell. Alignment changes the directional coherence and vector sum of these contributions without adding a separate propulsive term. The system remains actively driven when $\alpha=0$ and $v_0>0$. The zero-mean Gaussian velocity noise obeys $\langle\zeta_{am}(t)\zeta_{bn}(t')\rangle=2\mu k_BT_{eff}\delta_{ab}\delta_{mn}\delta(t-t')$. Here, $T_{eff}$ sets the amplitude of unresolved active fluctuations and is specified independently of the explicit active speed strength $v_0$. Polarity evolves on the unit sphere through alignment with neighboring cells, separation from close neighbors, inward steering near the spheroid contour, and rotational diffusion. Mechanical shape orientation is a separate observable and is not identified with polarity.

Activity in vertex models can also be represented by fluctuating junctional tensions, which generate equal and opposite forces at the ends of each edge \cite{keta2025long}, or by strain-dependent tension remodeling coupled to mechanical memory dissipation \cite{perezverdugo2023tension}. Here, each cell supplies directional active driving to its incident vertices, and neighboring cells can align their active directions. The independent vertex noise $\boldsymbol{\zeta}_a$ represents additional unresolved fluctuations and is distinct from fluctuating junctional tension.

This minimal framework combines cell deformation and neighbor exchange with active speed and local alignment steering under confinement. The three main control parameters are tissue state $s_0$, active speed $v_0$, and the alignment-to-rotational-diffusion ratio $\alpha/D_r$. We first vary activity with no alignment, then examine coherent motion in a separate high alignment regime. Compression is applied along $z$ by two frictionless plates, while contour steering confines motility. Geometry, numerical updates, loading protocols, and observable definitions are specified in Methods.

\begin{figure*}[htbp]
\centering
\includegraphics[width=\textwidth]{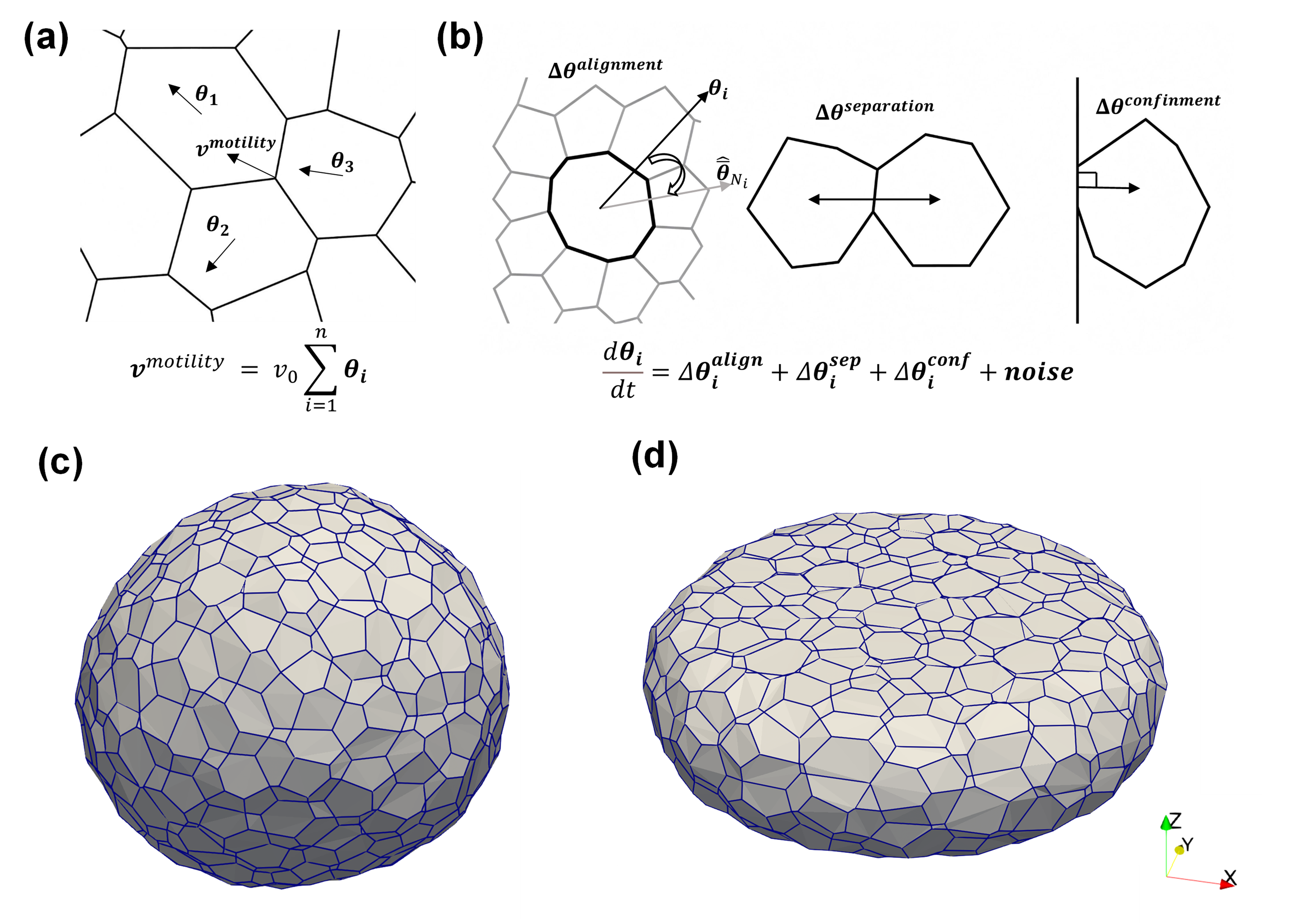}
\caption{{\it Three-dimensional active vertex model and spheroid geometry.}
(a) Each cell carries a unit active force direction $\boldsymbol{\theta}_i$
and contributes an active velocity $v_0\boldsymbol{\theta}_i$ to each incident vertex.
The active velocity imparted to a shared vertex is the sum of these contributions from the $n$ incident cells,
$\mathbf{v}^{\mathrm{motility}}=v_0\sum_{i=1}^{n}\boldsymbol{\theta}_i$,
while the actual cell velocity also includes mechanical and noise contributions.
(b) The three steering terms that reorient the polarity: alignment toward the mean
polarity of neighboring cells (gray), separation away from close neighbors, and
confinement, which rotates the polarity toward the inward normal $\hat{\mathbf{n}}_{\mathrm{in}}$
when the cell approaches the spheroid contour.
Rendered solid-like polyhedral cell networks of (c) the uncompressed spherical spheroid and
(d) the spheroid after uniaxial compression to $60\%$ axial strain.}
\label{fig:vertex_model_spheroid_geometry}
\end{figure*}

\section{Zero alignment: activity-dependent responses to compression} \label{sec:lowalign}

We first examine how compression changes the magnitude of disordered cell motion. To separate this response from coherent rotation, we switched alignment off ($\alpha=0$) and varied the active speed $v_0$ and compression strain in laterally confined solid-like ($s_0=5.0$) and fluid-like ($s_0=5.6$) spheroids. For each run, we measured the difference between mean in-plane cell speeds in the pre-compression and late post-compression sampling windows, $\Delta|v_{xy}|=\langle|v_{xy}|\rangle_{\rm post}-\langle|v_{xy}|\rangle_{\rm pre}$, with each bracket averaging over cells and sampled times. Motion along the compression axis is directly suppressed by the plates, so this measure isolates the transverse response. The activity range and replicate counts are given in the caption of Fig.~\ref{fig:zero_alignment_compression_response}.

\subsection{Activity reverses the fluid-like response to compression}

\begin{figure*}[htbp]
\centering
\includegraphics[width=0.9\textwidth]{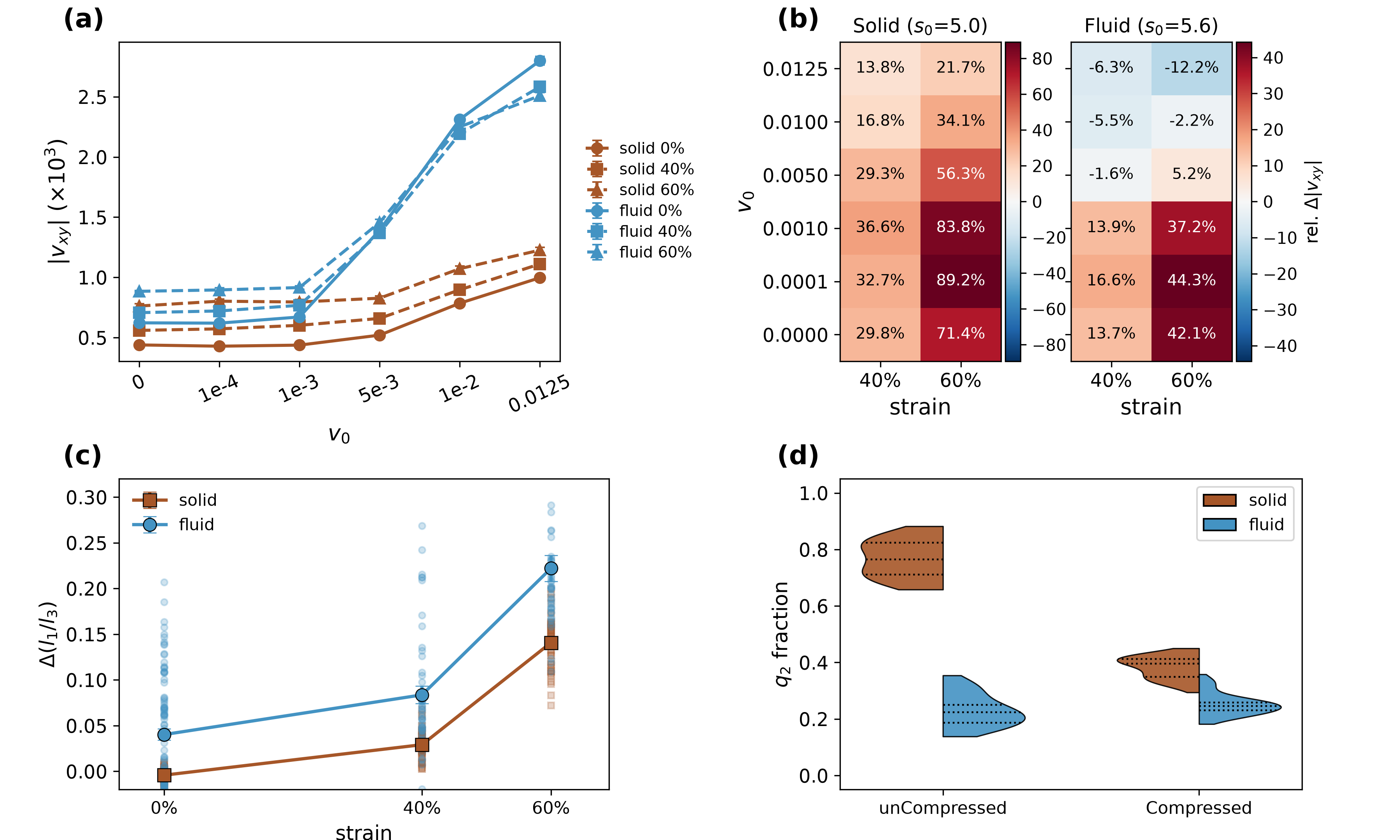}
\caption{{\it Compression response of flows with no alignment.} Laterally confined spheroids contain $250$ cells and are solid-like ($s_0=5.0$, brown) or fluid-like ($s_0=5.6$, blue), with active speed $v_0\in[0,1.25\times10^{-2}]$ and compression strains of $40\%$ and $60\%$. There are eight replicates per condition except for the highest-activity fluid at $60\%$ strain, which has five. Speed changes compare pre-compression and late post-compression time-window averages. (a) Mean in-plane cell speed $|v_{xy}|$ versus $v_0$ before compression and at $40\%$ and $60\%$ strain (mean $\pm$ SEM over replicates). Solid spheroids speed up in every group, with larger gains at $60\%$ strain; the absolute gain peaks at intermediate activity. Fluid spheroids speed up at low $v_0$ and slow down at high $v_0$. (b) Mean relative speed change, $\Delta|v_{xy}|/\langle|v_{xy}|\rangle_{\rm pre}$, across activity and strain. (c) Per-cell aspect-ratio change $\Delta(l_1/l_3)$ versus strain, pooled across activity conditions (large symbols: mean $\pm$ SEM over runs; small points: individual runs). (d) Split-violin distributions of per-run $q_2$ immediately before compression and at the final recorded moment, for $v_0=0.0125$ and $40\%$ strain (eight runs per tissue; dotted marks show the quartiles).}
\label{fig:zero_alignment_compression_response}
\end{figure*}

Solid spheroids speed up in every tested activity--strain group (Fig.~\ref{fig:zero_alignment_compression_response}a). The gain is larger at $60\%$ than at $40\%$ strain and reaches $70$--$90\%$ of the pre-compression speed at $60\%$ strain for $v_0\le10^{-3}$ (Fig.~\ref{fig:zero_alignment_compression_response}b). The absolute speed gain peaks at intermediate activity and declines at the highest $v_0$, while remaining positive. The enhancement is also present at $v_0=0$.

Fluid spheroids instead change the sign of their response as activity increases. They speed up at low $v_0$ and slow down at high $v_0$ (Fig.~\ref{fig:zero_alignment_compression_response}a--b). At $v_0=0.0125$, the relative changes are approximately $-6.3\%$ and $-12.2\%$ at $40\%$ and $60\%$ strain, respectively; the latter corresponds to a transverse speed decrease of $3.51\times10^{-4}$. The crossover also depends on strain: at $v_0=0.005$, the response is slightly negative at $40\%$ strain but positive at $60\%$. Thus, the solid-like or fluid-like classification alone does not fix the sign of the compression response: even at a fixed target shape index, changing activity can reverse the fluid-like response.

These speed changes occur without comparable coherent rotation over the observation window. The zero-alignment runs accumulate much smaller net angular displacements than the rotating high-alignment solids (Fig.~\ref{fig:zero_alignment_angular_displacement}). Their projected velocity--shape association stays close to the isotropic two-dimensional expectation $2/\pi$, with a modest increase in the most active fluids (Fig.~\ref{fig:zero_alignment_velocity_shape_correlation}). Compression also increases the cell aspect ratio, $l_1/l_3$; averaged across activity conditions, fluid cells deform more than solid cells (Fig.~\ref{fig:zero_alignment_compression_response}c). We next examine the contact states associated with these contrasting transverse responses.

\subsection{Contact persistence and the transverse cell speed response}

Here $q_2$ measures contact persistence relative to a reference neighbor set, as defined in Methods.
We compare endpoint values immediately before loading and at the final recorded time, defining the neighbor-retention loss as $-\Delta q_2=q_2^{\rm pre}-q_2^{\rm post}$. At the representative high-activity condition $v_0=0.0125$ and $40\%$ strain, the solid starts with substantially higher contact retention and loses much of it after compression, whereas the fluid has low retention both before and after loading (Fig.~\ref{fig:zero_alignment_compression_response}d).

Across the activity sweep, the fluid's pre-compression contact retention decreases steeply as $v_0$ increases, while the solid retains more neighbors throughout (Fig.~\ref{fig:activity_neighbor_retention}a). At $v_0=0$ and $60\%$ strain, the fluid has $q_2^{\rm pre}\approx0.84$, and compression is accompanied by a retention loss of approximately $0.60$ (Fig.~\ref{fig:activity_neighbor_retention}b) and a transverse speed increase of $2.62\times10^{-4}$ (Fig.~\ref{fig:zero_alignment_compression_response}a--b). At $v_0=0.0125$, its pre-compression values are only $0.230$ and $0.196$ for the $40\%$- and $60\%$-strain groups, respectively, and the corresponding endpoint losses are $-0.022$ and $0.055$. Thus, the low-activity fluid undergoes a large reduction in retained neighbor sets, whereas the high-activity fluid already has low retention before loading and exhibits little additional endpoint change.

The solid's pre-compression retention also decreases at higher activity, but remains above the fluid's (Fig.~\ref{fig:activity_neighbor_retention}a). Its retention loss remains substantial at both strains, even as its speed gain declines at the highest activities (Figs.~\ref{fig:activity_neighbor_retention}b and \ref{fig:zero_alignment_compression_response}b). These comparisons show that the same imposed strain acts on contact networks with different persistence across tissue states and activities.

\begin{figure}[htbp]
\centering
\includegraphics[width=\columnwidth]{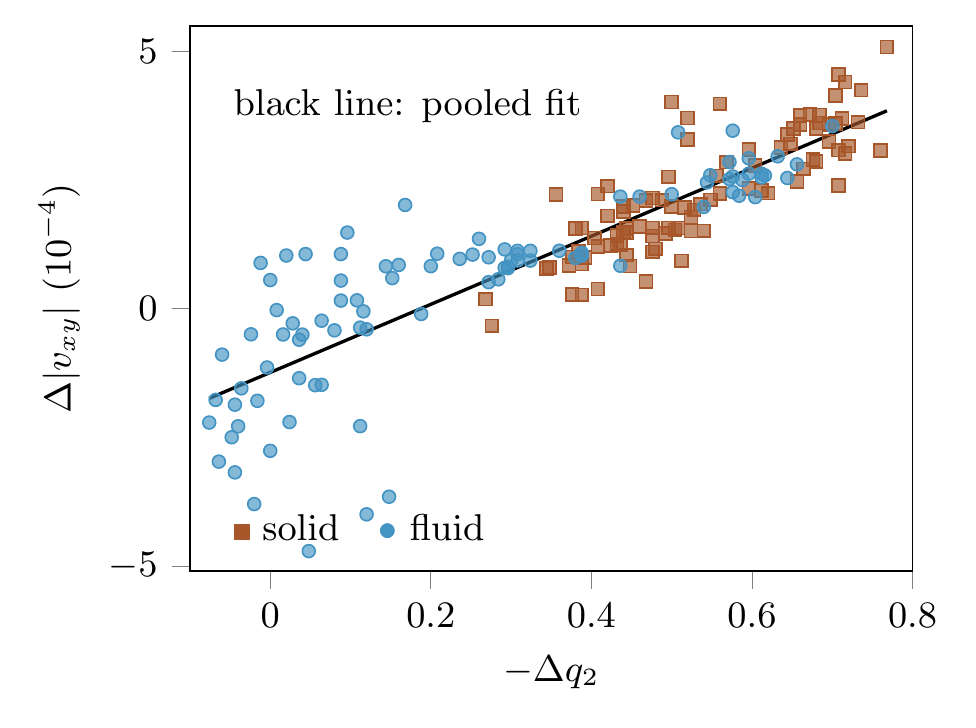}
\caption{{\it A relation between transverse speed and contact retention.} Change in mean in-plane cell speed versus endpoint neighbor-retention loss, $-\Delta q_2=q_2^{\rm pre}-q_2^{\rm post}$, with one point per run. The pooled fit is $\Delta|v_{xy}|=-1.24\times10^{-4}+(6.62\pm0.55)\times10^{-4}(-\Delta q_2)$, where the quoted slope uncertainty is one cluster-robust standard error.}
\label{fig:speed_neighbor_loss}
\end{figure}

Pooling the runs reveals a common association between the contact response and transverse motion: greater endpoint neighbor-retention loss is associated with a larger speed change (Fig.~\ref{fig:speed_neighbor_loss}). The fitted relation is $\Delta|v_{xy}|=-1.24\times10^{-4}+(6.62\pm0.55)\times10^{-4}(-\Delta q_2)$, with one cluster-robust standard error quoted for the slope. With its negative intercept, the fitted relation places acceleration and slowing on a common empirical trend, linking the activity-dependent speed response to changes in neighbor retention. We next examine how compression changes the local mechanical cost of rearrangement.

\subsection{Compression reshapes local rearrangement barriers}

To examine the mechanical cost of local rearrangement, we first considered an ordered geometric reference. Regular truncated octahedra tile three-dimensional space. For a single such cell, shortening one edge at fixed unit volume leads to a limiting shape index of approximately $s\simeq5.288$. This single-cell construction does not constrain the neighboring cells to deform compatibly. We therefore also considered a five-cell cluster of regular truncated octahedra and minimized its total vertex-model energy as the central edge shortened, retaining the coupled geometry of the neighboring cells. Its barrier vanishes near $s_0\simeq5.3$, still below the transition of the disordered system.

In two dimensions, the regular-polygon construction associated with a T1 rearrangement gives a useful estimate of the rigidity threshold~\cite{bi2015density}. The ordered constructions considered here underestimate the three-dimensional threshold even after local neighbor coupling is included. This discrepancy motivates examining disordered local environments, whose cell shapes and surrounding geometries differ from the regular truncated-octahedral reference.

\begin{figure*}[htbp]
\centering
\includegraphics[width=\textwidth]{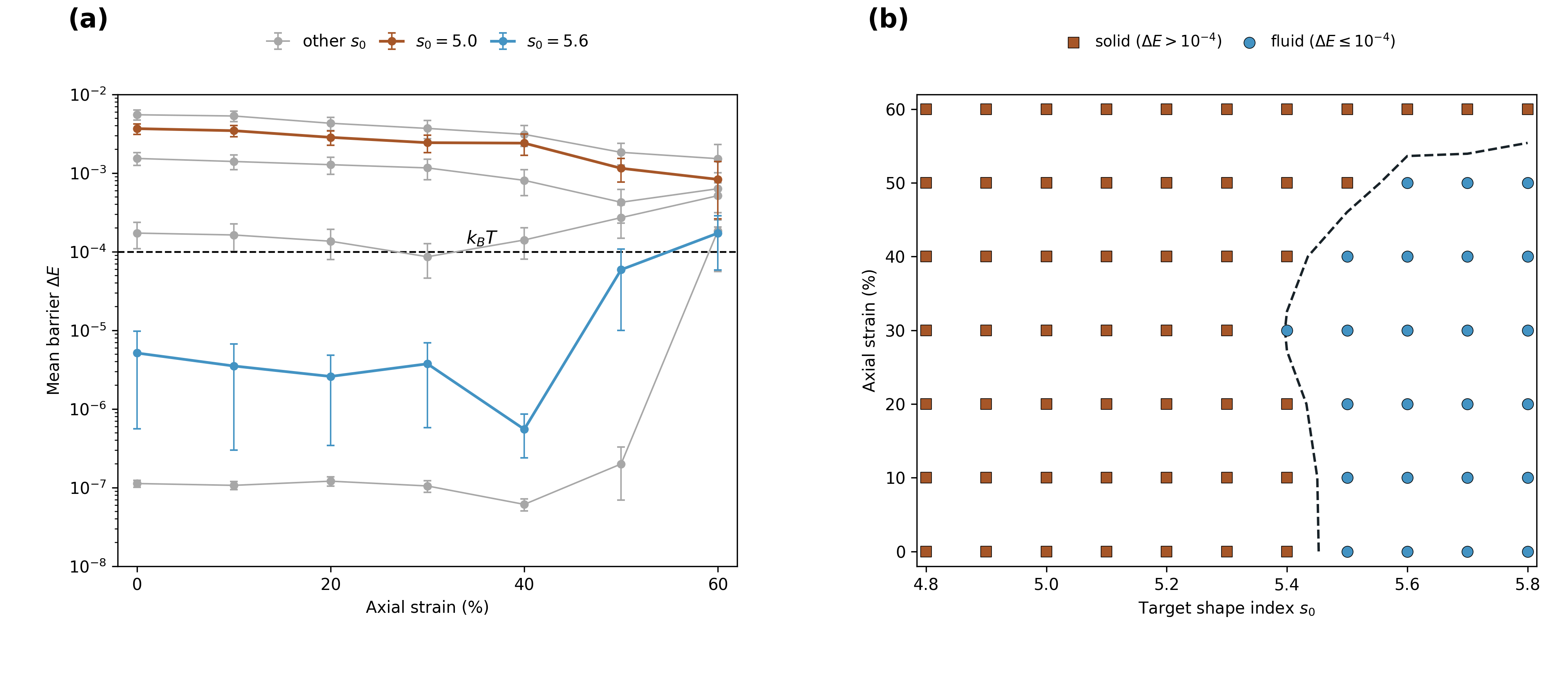}
\caption{{\it Local rearrangement barriers and mean-barrier classification in disordered five-cell clusters.}  (a) Mean energy barrier along the prescribed control-edge shrinkage path versus compression strain (mean $\pm$ SEM over topologies). Colored curves show $s_0=5.0$ and $5.6$; gray curves show $s_0=4.8$, $5.2$, $5.4$, and $5.8$ from the same sweep. The dashed line marks $\Delta E=10^{-4}$, equal to the effective noise energy scale $k_BT_{eff}$ used in the zero-alignment full-spheroid simulations. Compression along $z$ lowers the mean for $s_0=5.0$; for $s_0=5.6$, the mean falls from $30\%$ to $40\%$ strain and rises at higher strain. (b) Mean-barrier classification of the five-cell model in the $s_0$--$z$-strain plane. Each point is classified using the mean over twenty independent topologies and the threshold $\Delta E=10^{-4}$: brown squares denote a mean above the threshold (labelled solid), and blue circles denote a mean at or below it (labelled fluid). At zero strain the boundary lies between $s_0=5.4$ and $5.5$. It is weakly nonmonotonic at lower strains, shifts towards higher $s_0$ at high strain, and leaves the scanned range between $50\%$ and $60\%$ strain. At $60\%$ strain, the mean barrier therefore exceeds the threshold for every scanned $s_0$. This classification specifies the mean-barrier category of the passive five-cell model. The distribution of individual barriers is shown in Fig.~\ref{fig:t1_barrier_distribution}. Note that both panels use twenty independent Voronoi topologies subjected to $z$-directed pre-compression in the aligned cluster frame.}
\label{fig:t1_barrier_phase_diagram}
\end{figure*}

We next used five-cell clusters cut from random bulk Voronoi tissues. In three dimensions, an edge-to-triangular-face transition provides a local neighbor-exchange pathway. We measure the edge-collapse side of this pathway by shortening one interior control edge in discrete steps and minimizing the energy of the remaining free coordinates under the retained plate constraints, with topology fixed throughout, and define the barrier as $\Delta E=E_{\max}-E_{\rm init}$. The constrained minimizations use multi-start L-BFGS-B, and results are averaged over twenty independent Voronoi topologies. These passive clusters have no surface tension, activity, or alignment; their preparation and parameters are specified in Methods. Figure~\ref{fig:t1_barrier_phase_diagram}a shows the strain dependence of their mean barriers, and Fig.~\ref{fig:t1_barrier_phase_diagram}b summarizes the corresponding threshold classification across $s_0$ and strain. Without compression, the mean barrier falls with increasing $s_0$ and crosses the chosen $10^{-4}$ threshold between $s_0=5.4$ and $5.5$ (Fig.~\ref{fig:t1_barrier_phase_diagram}b). The solid-like value $s_0=5.0$ lies on the high-barrier side, whereas the fluid-like value $s_0=5.6$ has a mean below the effective noise energy scale $k_BT_{eff}=10^{-4}$ used in the zero-alignment spheroid simulations.

For compression along $z$ in the aligned cluster frame, pre-compression lowers the solid's mean barrier (Fig.~\ref{fig:t1_barrier_phase_diagram}a), showing that compression can lower the cost of this prescribed local rearrangement path. The fluid's mean responds differently: it falls from $3.74\times10^{-6}$ at $30\%$ strain to $5.50\times10^{-7}$ at $40\%$, then rises to $1.72\times10^{-4}$ at $60\%$. The increase therefore occurs at high strain rather than between $30\%$ and $40\%$. Across the $s_0$ sweep, the mean-barrier boundary is weakly nonmonotonic at lower strains and shifts towards higher $s_0$ at high strain, leaving the scanned range between $50\%$ and $60\%$ (Fig.~\ref{fig:t1_barrier_phase_diagram}b).

The response also depends on loading direction in the aligned cluster frame. With compression along $x$, the mean barriers at high strain exceed their uncompressed values for both representative $s_0$ values (Fig.~\ref{fig:directional_barriers_si}). For the fluid-like cluster, the high-strain mean under $x$ loading exceeds that under $z$ loading. The frame is defined by the initial middle-cell plane, so these two protocols compare loading within and normal to that plane; they do not impose an exact angle between the load and the relaxed control edge.

The distributions clarify what these means represent (Fig.~\ref{fig:t1_barrier_distribution}). The high-strain rise in the fluid's mean under $z$ loading is driven by a minority of high-barrier topologies; its median is zero at $60\%$ strain. At that strain, approximately $80\%$ of realizations lie at or below $\Delta E=10^{-4}$ for each of the two representative $s_0$ values. The fluid's median under $x$ loading also remains zero despite its larger mean (Fig.~\ref{fig:directional_barriers_si}). At high strain, the fluid-like clusters therefore combine a high-barrier tail with a majority of low-barrier paths. The distributions reveal a heterogeneous mechanical response that is not represented by the mean alone.

The activity-dependent speed reversal follows a common empirical relation with changes in neighbor retention (Fig.~\ref{fig:speed_neighbor_loss}). Low-activity fluids have high contact persistence before loading and, like solids, exhibit substantial retention loss and enhanced transverse motion after compression. High-activity fluids begin with much lower retention, show little additional endpoint loss, and slow down. Under $z$-directed compression, the reduced mean barrier in solid-like clusters is consistent with the enhanced neighbor loss and transverse motion observed in solid-like spheroids. Fluid-like clusters exhibit a heterogeneous barrier response, with a high-strain increase in the mean coexisting with many low-barrier paths. The local calculations thus identify how compression modifies mechanical resistance along prescribed rearrangement paths.

Throughout this sweep alignment was held at zero, and compression primarily changed the magnitude of disordered transverse motion over the observation window. We next examine coherent flow in a separate high-alignment series. That series uses different target shape indices, temperature, loading rate, and observation duration (Methods), so the comparison between the two series describes distinct parameter regimes.

\section{High values of alignment: Collective rotation} \label{sec:rotation}

\subsection{Compression induces coherent rotation selectively in solid-like spheroids} \label{sec:phenomenon}

We examine how compression affects collective motion at strong cell--cell alignment using solid-like spheroids at $s_0=4.8$ and fluid-like spheroids at $s_0=5.8$. At the representative parameters $\alpha=0.3$ and $v_0=0.005$, the spheroids were subjected to target axial strains of $0\%$, $30\%$, or $60\%$. After loading, the plates were held fixed and cell motion in the $xy$ plane was followed over time. Tissue state and strain are compared within this common high-alignment protocol.

\begin{figure*}[htbp]
\centering
\includegraphics[width=\textwidth]{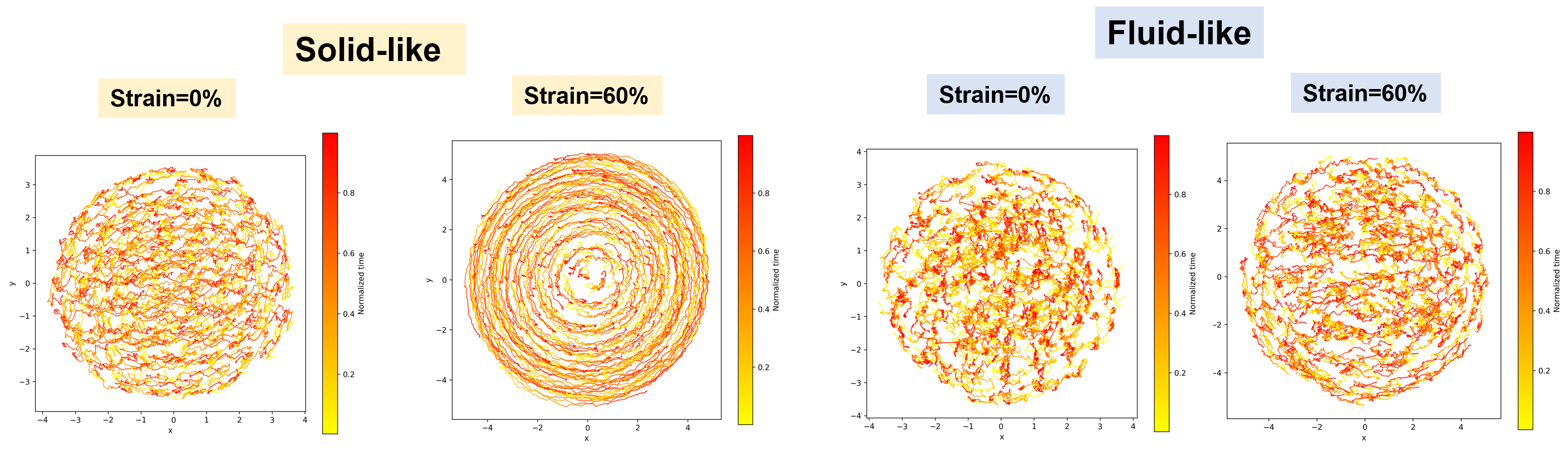}
\caption{{\it Cell displacement trajectories in the $xy$ plane.} From left to right: solid-like spheroid at $0\%$ and $60\%$ strain, followed by fluid-like spheroid at $0\%$ and $60\%$ strain. The compressed solid exhibits coordinated circumferential motion, whereas the other conditions show no comparable coherent circulation. All conditions use the same initial structure, sampling window, and analysis parameters, with alignment strength $\alpha=0.3$ and active speed $v_0=0.005$.}
\label{fig:high_alignment_cell_trajectories}
\end{figure*}

The trajectories reveal distinct collective responses in the two tissue states. Without compression, neither exhibits comparable coordinated circulation. At $60\%$ strain, the solid develops circumferential cell motion, whereas the fluid remains disordered under the same loading and activity conditions (Fig.~\ref{fig:high_alignment_cell_trajectories}). The four panels show, from left to right, the uncompressed and compressed solid followed by the uncompressed and compressed fluid.

\begin{figure*}[htbp]
\centering
\includegraphics[width=\textwidth]{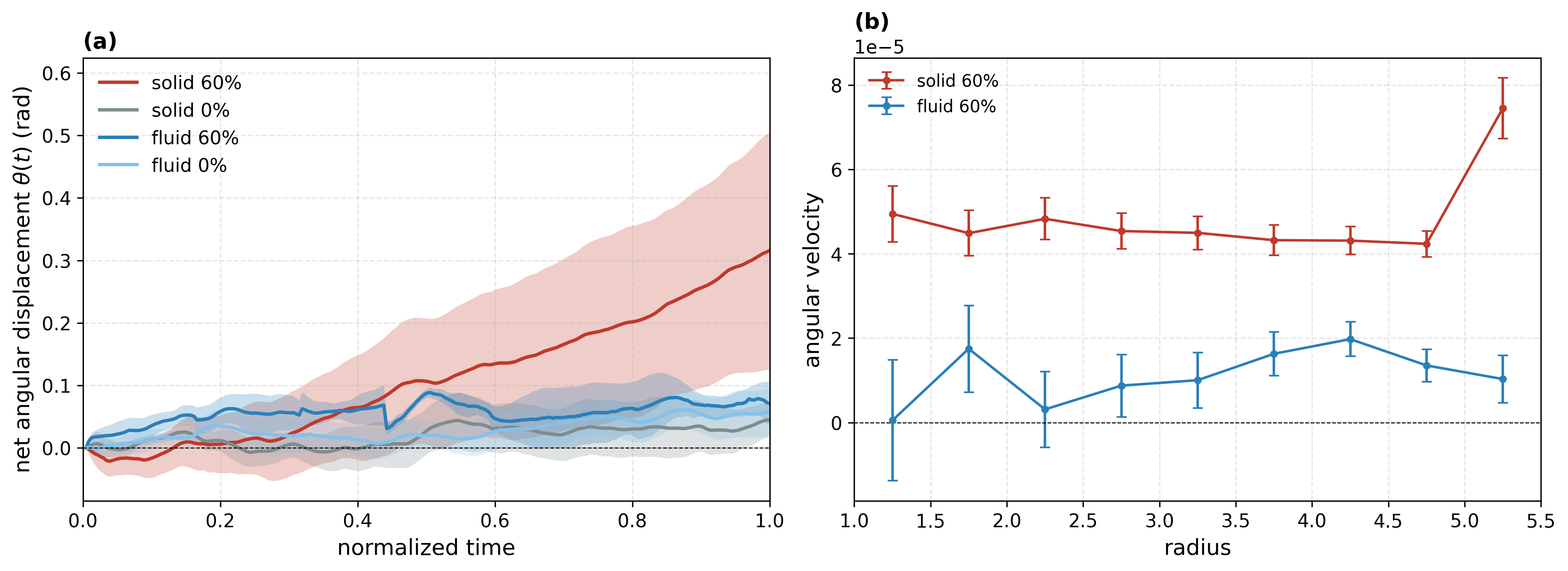}
\caption{{\it Quantitative evidence for coherent rotation in compressed solid-like spheroids.} (a) Time-integrated signed angular displacement $\theta(t)$ for solid and fluid spheroids at $0\%$ and $60\%$ strain, plotted against time normalized to $[0,1]$ over the sampled window. Each run is multiplied by $\operatorname{sgn}(\theta(t_{\rm final}))$ before averaging to account for spontaneous clockwise or counterclockwise drift. Curves show mean $\pm$ SEM over replicates and are drawn while at least two replicates remain. The compressed solid accumulates the largest mean angular displacement. (b) Sign-normalized angular velocity versus radius at $60\%$ strain (mean $\pm$ SEM over sampled times). The solid has same-sign angular velocity throughout, with a relatively uniform interior and a larger value near the periphery; the fluid has a smaller angular-velocity profile.}
\label{fig:collective_rotation_dynamics}
\end{figure*}

To quantify this motion, we integrate the aggregate's signed angular velocity over time. Clockwise and counterclockwise drift occur across replicates, so each trajectory is normalized by the sign of its final angular displacement before averaging. The mean accumulated displacement is substantially larger in the solid at $60\%$ strain than in the uncompressed solid or either fluid condition (Fig.~\ref{fig:collective_rotation_dynamics}a). The motion remains slow, amounting to a fraction of one revolution over the observation window.

The radial angular-velocity profile provides a complementary measure of coordination. The compressed solid exhibits coherent, approximately rigid-body-like rotation: its sign-normalized angular velocity has the same sign throughout, with a nearly uniform interior and a faster periphery (Fig.~\ref{fig:collective_rotation_dynamics}b). The fluid response is smaller. Together, the trajectories and angular measurements establish coherent rotation in the compressed solid under these conditions.

The solid's response also depends on the applied strain. Across the tested $0\%$, $30\%$, and $60\%$ conditions, the mean final angular displacement increases, with the intermediate strain giving an intermediate response (Fig.~\ref{fig:strain_dependent_rotation_alignment}a,c). The corresponding mean peripheral tangential angle decreases (Fig.~\ref{fig:strain_dependent_rotation_alignment}c). Across the tested strains, stronger compression therefore increases the mean accumulated rotation and strengthens peripheral tangential organization. We next examine why the solid and fluid respond differently by comparing their contact persistence, deformation, and mechanical organization.

\subsection{Mechanistic interpretation of coherent rotation} \label{sec:mechanism}

To explain compression-induced coherent rotation, we examine two aspects of the response: the spatial organization of tangential cell motion and the mechanical response to azimuthal deformation. We first relate contact persistence and peripheral cell organization to the observed motion. We then use controlled twist perturbations to determine how compression and activity change the solid's mechanical response. These analyses motivate a feedback model in which azimuthal deformation and coordinated active motion reinforce one another.

\subsubsection*{Contact persistence and organization of tangential motion}

\begin{figure*}[htbp]
\centering
\includegraphics[width=\textwidth]{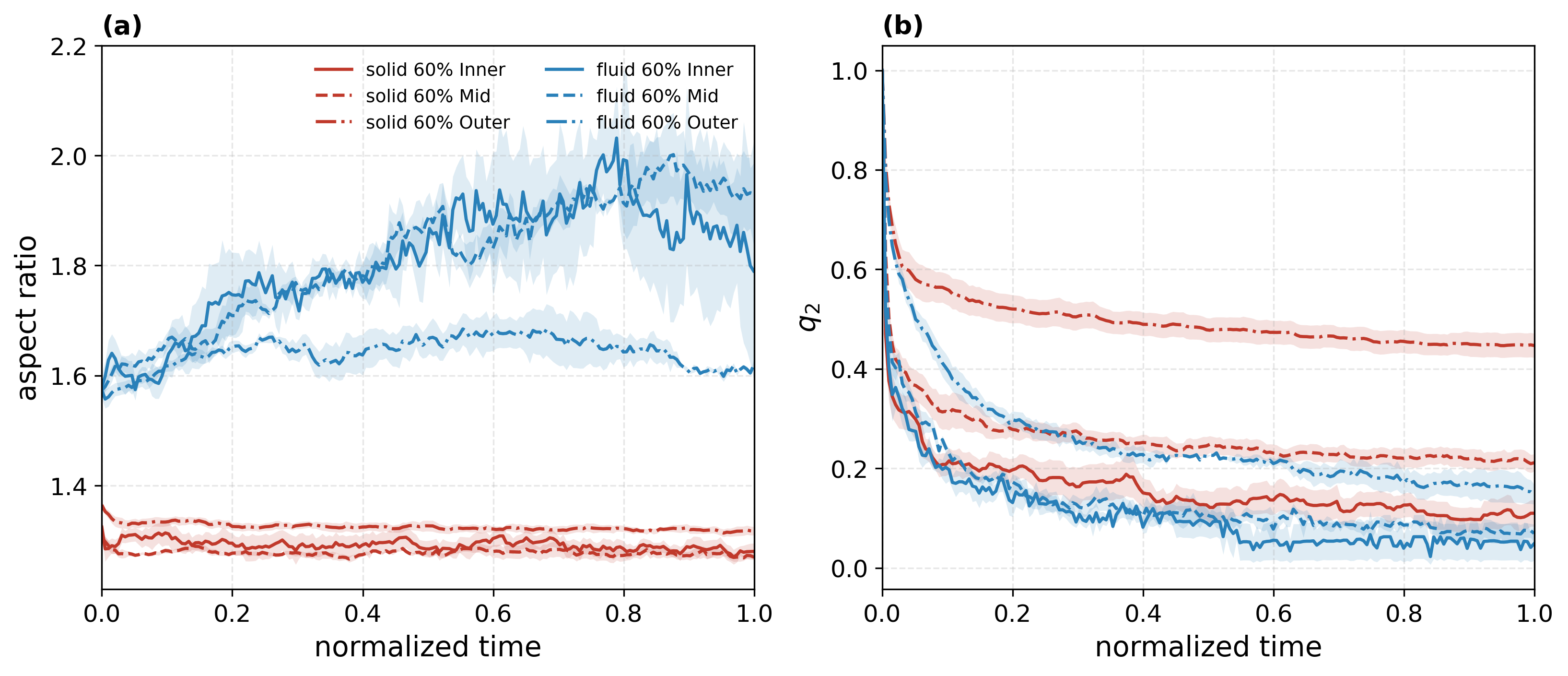}
\caption{{\it Cell deformation and neighbor persistence at $60\%$ strain.} (a) Aspect ratio versus time. Solid spheroids have their largest aspect ratio in the outer layer; fluid spheroids do not show the same peripheral maximum. (b) Contact-persistence index $q_2$ versus time: a cell contributes one if every neighbor in its reference set is still present. After loading, the solid shows an initial drop followed by slower decay; the fluid has a lower contact-persistence index. Inner, Mid, and Outer denote the three radial bins. The mean $\pm$ SEM over replicates is shown with curves drawn as long as least two replicates remain; time normalized to $[0,1]$ over the sampled window. }
\label{fig:cell_elongation_neighbor_stability}
\end{figure*}

The uncompressed solid maintains relatively high contact persistence, with $q_2\approx0.77$ at the end of the observation window (Fig.~\ref{fig:neighbor_retention_relaxation}). After loading, the compressed solid shows an initial drop in $q_2$, followed by slower decay. Thus, the compressed solid exhibits coherent rotation while maintaining higher contact persistence than the fluid (Fig.~\ref{fig:cell_elongation_neighbor_stability}b).

Cell deformation is measured by the aspect ratio of the projected long and short axes. At $60\%$ strain, the solid's outer cells have the largest aspect ratio, whereas its inner and middle layers are less elongated (Fig.~\ref{fig:cell_elongation_neighbor_stability}a). Fluid spheroids also show radial differences in shape, but do not exhibit the same peripheral maximum. We next examine the stress distribution and cell-axis orientation associated with this peripheral deformation.

\begin{figure*}[htbp]
\centering
\includegraphics[width=\textwidth]{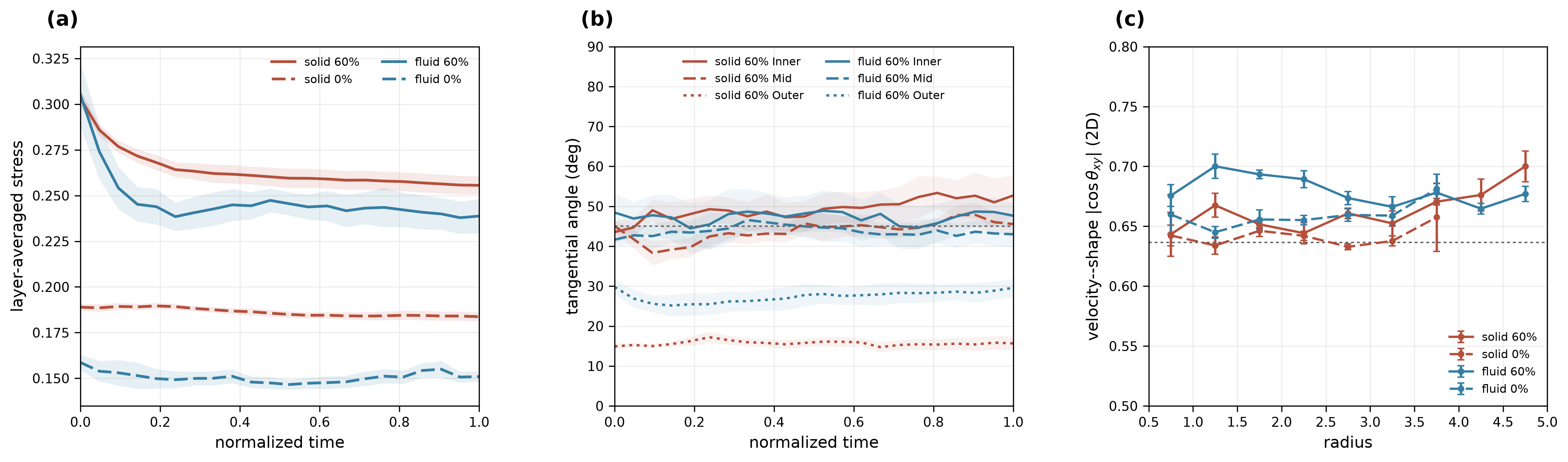}
\caption{{\it Vortex mechanism dynamics at early times.}
(a) Layer-averaged stress versus time: compression elevates the solid's stress
well above its uncompressed value.  (b) Tangential angle versus time at 60\%
strain at three radial bins: outer cells align most strongly with the
tangential direction (smallest angle) in both tissues, with the solid showing
the deepest alignment; the dotted line marks the random expectation
($45^\circ$).  (c) Mean absolute cosine between successive-snapshot cell-centroid displacement and the longest shape axis, both projected onto the $xy$ plane, versus projected radius from the spheroid centroid. The dotted line marks the isotropic two-dimensional random expectation $2/\pi$.
}
\label{fig:stress_shape_alignment_dynamics}
\end{figure*}

Compression raises the solid's layer-averaged scalar stress above its uncompressed value (Fig.~\ref{fig:stress_shape_alignment_dynamics}a). During the early window $579\le t\le1000$ at $60\%$ strain, its outer-layer mean is $0.291$, compared with $0.226$ and $0.220$ in the inner and middle layers (Fig.~\ref{fig:radial_stress_early_si}a). The fluid has a different radial profile: its outer-layer scalar stress lies below the inner and middle values (Fig.~\ref{fig:radial_stress_early_si}b).

We compare these profiles with the tangential angle between the projected cell long axis and the local circumferential direction. In the compressed solid, peripheral cells are more tangentially oriented than inner cells (Fig.~\ref{fig:stress_shape_alignment_dynamics}b). The strain comparison also shows a smaller mean final peripheral tangential angle at larger compression (Fig.~\ref{fig:strain_dependent_rotation_alignment}c). Fluid spheroids develop peripheral tangential orientation as well, although it is weaker than in the solid. The compressed solid exhibits a distinct peripheral organization: its outer layer combines the highest scalar stress with the strongest tangential cell orientation, whereas the fluid lacks this peripheral stress maximum. 

To connect cell organization with motion, we compute the mean absolute cosine, $\langle|\cos\psi_{xy}|\rangle$, between successive-snapshot centroid displacement and the longest cell-shape axis after both are projected onto the $xy$ plane. Values near unity indicate motion along the unsigned projected axis in either direction; independent isotropic directions in two dimensions give the reference value $2/\pi$.

At $60\%$ strain, the solid's mean association increases overall toward the periphery, from approximately $0.64$ in the innermost bin to $0.70$ at the outermost measured radius, with fluctuations between bins (Fig.~\ref{fig:stress_shape_alignment_dynamics}c). The outer value exceeds $2/\pi\approx0.637$, indicating a modest preference for peripheral motion along the projected long axis. The fluid also shows an association, with values of approximately $0.67$--$0.70$ across radial bins. Local shape--motion association is therefore present in both tissues, directing attention to how these local motions become coordinated around the spheroid.

The rotating solid combines a relatively persistent contact network with pronounced peripheral elongation and tangential orientation. The unsigned cell long axis accommodates both clockwise and counterclockwise motion. We hypothesize that the more persistent contact network in the solid helps polarity alignment coordinate tangential motion across cells and layers, organizing local motion into coherent circulation.

\subsubsection*{Compression softens differential azimuthal deformation}

At the same alignment strength, pronounced coherent rotation develops in the solid after compression. We therefore test how loading changes its mechanical response to azimuthal deformation. A controlled differential-twist perturbation probes the restoring forces associated with relative tangential displacements between radial layers. We impose a small displacement,
\[
\delta \mathbf r_i \propto
(-y_i,x_i,0)
\left(\frac{\rho_i}{R_{xy}}\right)^2,
\]
and remove the rigid-body rotational component. The radial dependence produces differential tangential shear between layers.

\begin{figure*}[htbp]
\centering
\includegraphics[width=0.96\textwidth]{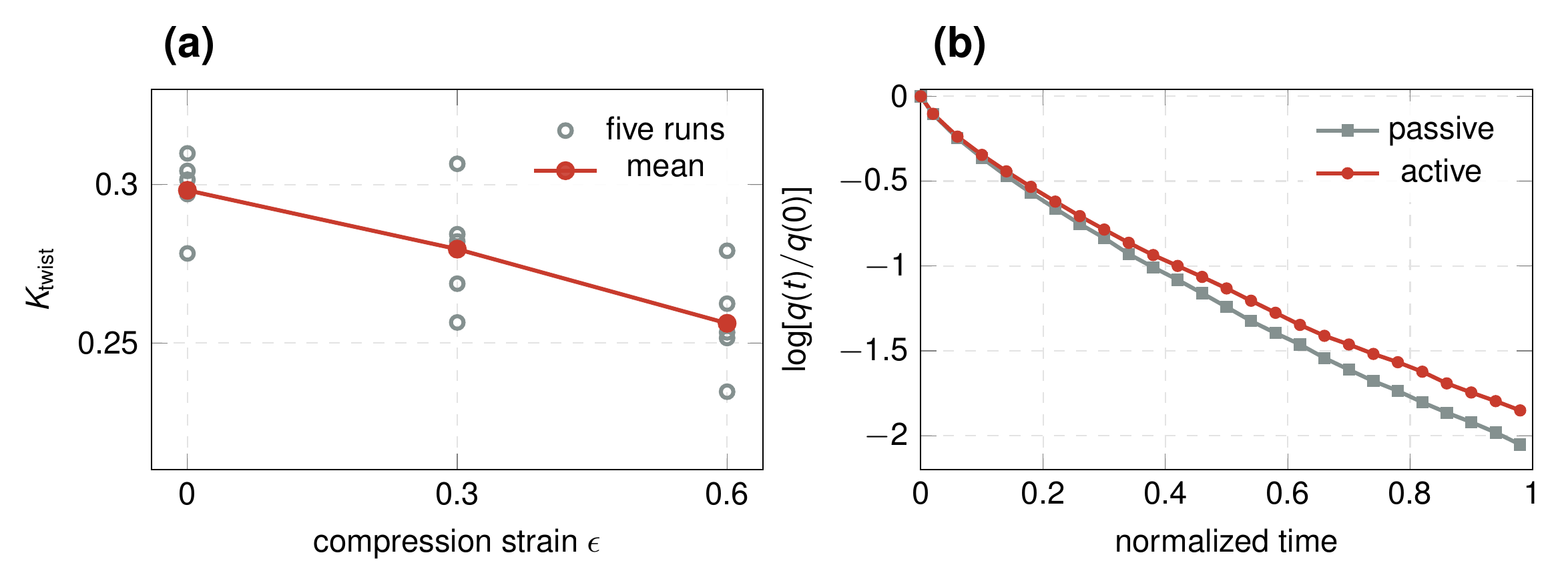}
\caption{{\it Compression-induced twist softening and slower relaxation with activity.} (a) Full projected force-response stiffness at saved time $t=799$ versus axial compression: open symbols show five matched topologies and filled symbols their means; the uncompressed baseline is not subtracted. (b) Relaxation of an imposed twist at 60\% compression, with and without activity, shown as the logarithm of the normalized projected twist amplitude. Time is normalized by the 50 model-time-unit probe duration after release; the last saved point is at $t=49$. Both perturbations relax, with the active response decaying more slowly. The relaxation curves are controlled probes from one saved geometry and a matched polarity initialization. Note that $s_0=4.8$. }
\label{fig:twist_response_high_alignment}
\end{figure*}

The effective twist stiffness, $K_{\rm twist}$, decreases systematically with compression, from approximately $0.298$ without compression to $0.256$ at $60\%$ strain (Fig.~\ref{fig:twist_response_high_alignment}a). This reduction demonstrates softening of differential azimuthal shear in the prestressed solid. After release at $60\%$ strain, the imposed twist decays both with and without activity; activity reduces the decay rate and prolongs the deformation (Fig.~\ref{fig:twist_response_high_alignment}b). The twist probes thus establish two mechanical effects: compression reduces the restoring response to differential azimuthal deformation, and activity prolongs its relaxation. These results motivate examining how this deformation can interact with the polarity coordination proposed above.

\subsubsection*{Deformation--polarization feedback: a heuristic model}

We connect the spatial and mechanical analyses through a proposed feedback between azimuthal deformation and polarity coordination. The differential-twist mode probes internal shear, whereas the trajectories show approximately rigid-body-like rotation. To connect these responses, we assume that azimuthal deformation promotes coherent azimuthal polarization and that the resulting active motion drives further deformation. We represent this feedback by coupling a differential-twist amplitude $A$ to a coherent azimuthal polarization $P$,
\[
\dot A=-\lambda_{\rm tw}(\epsilon)A+c_vv_0P,
\qquad
\dot P=-\lambda_pP+c_\alpha\alpha A.
\]
Here $\lambda_{\rm tw}$ is the passive twist-relaxation rate, $\lambda_p$ is the polarization-relaxation rate, and the coupling terms describe the hypothesized feedback between active motion, deformation, and alignment. For positive feedback, linear stability of $A=P=0$ gives
\[
\frac{c_vc_\alpha\alpha v_0}{\lambda_p}
\equiv\Lambda_{\rm act}
\gtrsim\lambda_{\rm tw}(\epsilon).
\]
This criterion identifies the active feedback required to overcome mechanical relaxation and produce growth of the coupled mode. Within the constant-mobility approximation, compression lowers the passive relaxation rate through the measured twist softening and therefore reduces the active feedback required for growth. With the other coefficients held fixed, the model predicts that increasing activity or alignment reduces the compression required for growth of the coupled mode. The numerical calibration and the loading-stress estimate are given in Methods. The contact and peripheral-organization comparisons identify structural differences between the rotating solid and the fluid, while the twist measurements and feedback model address the compression dependence within the solid.

The resulting mechanistic picture combines spatial coordination with a compression-dependent mechanical response. The rotating solid retains a more persistent contact network than the fluid and develops pronounced peripheral tangential organization. Compression also softens its differential azimuthal deformation, whose relaxation is slowed by activity. We propose that polarity alignment acting through persistent contacts couples coordinated tangential motion to this softened deformation mode, providing a route from compression to coherent circulation. Interestingly, compression-induced softening of the differential-twist mode does not necessarily imply an overall softening of the tissue. Tsubota et al. reported strain softening in three-dimensional vertex simulations, showing that multicellular structure can produce softening even without neighbor exchanges \cite{tsubota2025strain}. Our twist assay probes the restoring response to a small differential azimuthal deformation after compression, providing the mechanical component of the proposed mechanism for coherent rotation. Previous experiments have demonstrated compression stiffening in individual fibroblasts, with theoretical models identifying several microscopic mechanisms for this nonlinear response~\cite{gandikota2020loops}. Our results suggest that compression can modify distinct mechanical modes differently, potentially increasing resistance to axial deformation while reducing resistance to differential azimuthal motion.


\section{Discussion}

Our results show that one minimal physical framework combining cell mechanics, active forces, and neighbor alignment generates diverse changes in both the magnitude and organization of cell motion under compression. With alignment switched off, increasing activity reverses the fluid-like response from the transverse speeds of cells increasing to one in which the transverse speeds are decreasing, while solid-like spheroids always exhibit higher transverse speeds. Such trends can be understood, in part, via local, passive five-cell energy barrier calculations, in terms of the raising or lowering of the energy barrier to cell rearrangements. An extension of these local calculations would include the work performed by the cell-directed active forces along a rearrangement path. This would connect the compression-dependent mechanical cost to the magnitude and orientation of active driving, including its coordination by alignment.  Interestingly, in two dimensions, similar calculations using a regular geometry yielded an accurate prediction of the onset of the rigidity transition, while in three dimensions, the random geometry is needed for accuracy.  

In the rotating solid with each direction equally likely, coherent motion accompanies greater neighbor retention than in the fluid-like spheroid and tangential peripheral organization. At matched loading and material state, we hypothesize that stronger polarity coordination may promote coherent rotation, whereas reduced contact persistence may disrupt it. A related experimental link is provided by developing mammary epithelial structures, where transient compression restored early coherent rotation and E-cadherin blockade prevented compression-induced phenotypic reversion \cite{ricca2018transient}. These findings motivate testing junctional contributions to coordination.

Additional experiments provide direct context for these results. Compression-induced vortex motion has been observed in MCF10A spheroids but not in MDA-MB-231 spheroids under the same experimental protocol \cite{suh2025microfluidic}, and independent compression--recovery measurements establish differences in their elastic recovery and plastic deformation \cite{pandey2026viscoelastic}. The different motility responses reported under another compression protocol \cite{pandey2024mechanical} emphasize the need to account for loading conditions and active dynamics as well as tissue properties. Our simulations provide a framework for testing how activity and contact persistence contribute to these protocol-dependent responses. The coexistence of rotation and persistent contacts also connects our compression-driven result to active-solid rotation studied in pancreatic spheres \cite{tan2024emergent}.

In comparison with other computational models, rigid-body rotation emerges in confined two-dimensional cell monolayers through persistent active driving and mechanical coupling, even without explicit alignment between neighboring cells~\cite{li2014coherent}. Subsequently, it was shown that active vertex models can exhibit both global coherent rotation and more localized vortices~\cite{koride2018epithelial}. Thus far, our results in three dimensions demonstrate that coherent rotation depends on the interplay between tissue mechanical state, active driving, {\it and} directional alignment. In particular, the emergence of coherent rotation in solid-like but not fluid-like spheroids suggests that contact persistence and the energetic cost of cell rearrangements may play an important role in sustaining collective rotational motion, and thus we have not yet observed multiple vortices.

Okuda and Sato demonstrated collective cluster migration through surrounding tissue driven by polarized interfacial tension \cite{okuda2022polarized}. Here, cell activity contributes directly to shared-vertex motion, while neighboring cells dynamically coordinate their propulsion directions. We use this framework to examine how uniaxial compression changes internal motion in finite spheroids across solid-like and fluid-like regimes, revealing an activity-dependent reversal of the speed response and coherent rotation accompanied by persistent contacts.

Other recent work employed a three-dimensional active foam model to investigate spheroid fusion, demonstrating that cellular activity can promote tissue fluidization and complete fusion, while intercellular friction can suppress rearrangements and lead to arrested fusion~\cite{ongenae2025active}. Their model explicitly incorporates cortical viscosity, intercellular friction, and active protrusive forces, whereas our three-dimensional active vertex model provides a more minimal description based on cell volume and surface-area elasticity, active driving, and alignment of neighboring cells' active directions. Our results suggest that active collective motion need not require extensive cellular rearrangements: coherent rotation can occur in a relatively persistent contact network. This raises the possibility that, during spheroid fusion, activity and directional alignment could promote collective rotation without necessarily accelerating the rearrangements required for fusion. Conversely, compression-induced reductions in rearrangement barriers could facilitate fusion even in initially solid-like spheroids. Thus, extending our model to interacting spheroids could reveal how compression and directional alignment independently regulate collective motion and tissue fusion.

As for additional testable predictions, with cell alignment switched off, lowering activity can reverse the fluid-like response in terms of cell speed in the plane transverse to the uni-axial compression, with the crossover depending on compression strain. This prediction can be tested by varying activity and compression while independently monitoring the tissue's material state. Moreover, our high-alignment results predict coherent rotation accompanied by greater neighbor retention than in fluid-like spheroids under similar compression and activity. This could be tested by removing overall translation and rotation from cell trajectories and quantifying residual relative motion, neighbor retention, and rearrangement events. Moreover, active K-RAS can also induce coherent rotation of epithelial clusters and cysts \cite{hirata2018active}, with active K-RAS enhancing the activities. 

 As for next steps, it is known that distinct matrix-remodeling strategies can favor different collective motions: rotational migration is associated with global integrin-$\beta_1$-mediated remodeling, whereas invasive migration involves localized matrix-metalloproteinase activity \cite{ranamukhaarachchi2025global}.  It, this, would be of interest to understand the effects of the coupling between the spheroid and the ECM in the presence of compression. Moreover, we must go beyond the cell scale to the chromatin scale~\cite{zhang2026human} better understand how cell collective can adapt, or even be trained,~\cite{ameen2026training} given inputs from their environment. For now, our minimal model produces a diverse set of complex collective cellular motions with perhaps fixed parameters that can be ultimately identified with different cell types on short time scales and updatable parameters on longer timescales. 

\section*{Methods}

\subsection*{Geometry, boundary conditions, and initialization}

The spheroids contain 250 cells and use the energy in Eq.~\ref{eq:vertex_energy}, with $K_V=10$, $K_A=1$, and $\Gamma=1$. The volume term penalizes volume changes, whereas the area term represents effective cortical contractility and cell--cell adhesion. Lengths are reported in units of $V_0^{1/3}$, and the numerical mobility and Boltzmann constant are set to $\mu=k_B=1$.

To introduce a finite spheroid boundary, we construct the tissue by cutting out a spherical cluster from a bulk periodic confluent system, leaving empty space outside the cluster. Cells at the spheroid boundary therefore possess faces exposed to the exterior. These boundary faces contribute an additional interfacial energy, where \(\alpha\) labels cell faces and \(\delta_{\alpha,B}=1\) for boundary faces and \(0\) otherwise; \(\Gamma\) is the corresponding surface tension. In the simulations reported here we use \(\Gamma = 1\), which keeps the boundary intact.

The dynamical simulations start from an input cellular geometry and evolve for a prescribed pre-loading interval before the plates begin to move. Vertex motion, polarity updates, and permitted reconnections continue during this interval and during subsequent loading and observation.

All zero-alignment replicates share one input topology. The high-alignment simulations use five input topologies, each reused across tissue states and strains. No fixed random seed or initial-polarity assignment is specified in the saved configurations.

During the production trajectories, the configured active speed $v_0$ is applied from the start of the pre-loading evolution. The intervals preceding loading at $t=1500$ and $t=540$ therefore describe evolution under the specified activity and noise for the zero- and high-alignment series, respectively. The input topology files were exported after the spherical cut and used directly in the production simulations, without a separate passive relaxation.

\subsection*{Polarity dynamics and numerical integration}

Let $N_i$ denote the cells sharing a face with cell $i$. The vectors $\mathbf g_i^{\mathrm{align}}$, $\mathbf g_i^{\mathrm{sep}}$, and $\mathbf g_i^{\mathrm{conf}}$ denote deterministic steering contributions to the evolution of the cell's active direction, before projection onto its tangent plane; they are not forces acting on vertex positions. Before projection, the alignment and separation drives are
\[
\mathbf g_i^{\mathrm{align}}=\alpha\big(\widehat{\bar{\boldsymbol\theta}}_{N_i}-\boldsymbol\theta_i\big),
\]
\[
\mathbf g_i^{\mathrm{sep}}=
\frac{S_s}{|N_i^{\mathrm{close}}|}
\sum_{j\in N_i^{\mathrm{close}}}
\left(1-\frac{d_{ij}}{R_s}\right)
\frac{\mathbf c_i-\mathbf c_j}{d_{ij}} .
\]
Here $\mathbf c_i$ is the cell center, $d_{ij}=|\mathbf c_i-\mathbf c_j|$, the hat denotes normalization of the mean neighbor polarity, and $N_i^{\mathrm{close}}$ contains face neighbors with $10^{-3}<d_{ij}<R_s$. The separation drive is zero when this set is empty.

Motility confinement uses an ellipsoid fitted to the cell-center positions. Its center $\mathbf c$ is their mean, and its semiaxes are $a_m=\max_j|c_{jm}-c_m|$ for $m=x,y,z$. Define the normalized ellipsoidal radius $\rho_i=[\sum_m((c_{im}-c_m)/a_m)^2]^{1/2}$ and the outward unit normal $\hat{\mathbf n}_i$, parallel to $((c_{ix}-c_x)/a_x^2,(c_{iy}-c_y)/a_y^2,(c_{iz}-c_z)/a_z^2)$. The confinement drive is
\[
\mathbf g_i^{\mathrm{conf}}=-\beta[\rho_i-(1-b)]_+\hat{\mathbf n}_i,
\]
where $[u]_+=\max(u,0)$. This term reorients polarity inward within an outer shell of width $b$; it does not constrain vertex positions. The ellipsoid-update protocol is specified below.

The alignment and separation drives share a unit relaxation-time factor (\texttt{decay\_factor}$=1$), and the confinement drive is added separately. With this unit factor, the total deterministic steering vector is $\mathbf g_i=\mathbf g_i^{\mathrm{align}}+\mathbf g_i^{\mathrm{sep}}+\mathbf g_i^{\mathrm{conf}}$. At each step, the deterministic drive and a three-dimensional Gaussian increment are projected with $\mathbb P_i=\mathbb I-\boldsymbol\theta_i\boldsymbol\theta_i^{\mathsf T}$. The deterministic direction increment is $\Delta t\,\mathbb P_i\mathbf g_i$, and the projected rotational-noise increment has prefactor $\sqrt{2D_r\Delta t}$. The updated polarity is then normalized to unit length. Vertex coordinates are advanced by Euler--Maruyama integration of Eq.~\ref{eq:vertex_dynamics}, with translational Gaussian increments of prefactor $\sqrt{2\mu k_BT_{eff}\Delta t}$. The dynamics are overdamped and contain no inertia.

The supplied zero-alignment run configuration uses $\Delta t=0.01$, $T_{eff}=10^{-4}$, $D_r=10^{-5}$, $S_s=0.5$, $R_s=0.5$, $b=0.1$, and $\beta=2$. The simulation subtracts the mean mechanical vertex velocity in the $xy$ plane, and also along $z$ before loading, and recenters the real-vertex positions after each position update. Cell displacement is measured in this centered frame.

The high-alignment production runs instead use $T_{eff}=10^{-3}$, with the same $\Delta t=0.01$, $D_r=10^{-5}$, $S_s=0.5$, $R_s=0.5$, $b=0.1$, and $\beta=2$. Both series save frames every 20 model-time units.

\subsection*{Topological rearrangements}

Neighbor exchanges are implemented by three-dimensional I--H transformations~\cite{Okuda2013}. An I-to-H event replaces a short edge by a triangular face; an H-to-I event performs the reverse transformation. An edge is eligible for I-to-H conversion only if its length is below the reconnection threshold and it is not part of a triangular face. H-to-I conversion requires all three edges of a triangular face to be below the threshold, each belonging to only one triangular face, together with the local topology checks in the reconnection routine. A single short edge on a triangular face does not by itself trigger H-to-I conversion.

In the supplied zero-alignment configuration, the length threshold is $0.02$ and reconnections are checked every 10 integration steps, corresponding to $0.1$ model time units. These checks remain enabled during pre-loading evolution, compression, and subsequent observation.

The high-alignment production runs also use a threshold of $0.02$; reconnections are checked every $0.1$ model-time units during pre-loading, compression, and subsequent observation.

\subsection*{Loading protocols, parameter sets, and time calibration}

The spheroid is placed in an otherwise empty, non-periodic cubic box and compressed uniaxially along the \(z\)-axis between two parallel virtual plates.
At the start of loading, the plate half-gap is $H_0=\max_a|z_a|$ over real-cell vertices. Each plate moves toward the midplane at speed $u_{\rm p}$ until $H=H_0(1-\epsilon)$, where $\epsilon$ is the target axial strain, and is then held fixed. The full-spheroid code uses soft, frictionless plate forces: a vertex with $|z_a|>H$ experiences an inward axial force of magnitude $\min\{80(|z_a|-H),20\}$, while a vertex inside the gap feels no plate force. There is no tangential plate force or hard coordinate projection.

The zero-alignment configuration runs from $t=0$ to $2000$, starts loading at $t_{\rm load}=1500$, and uses $u_{\rm p}=4$ for each plate. The activity sweep uses $v_0=0,10^{-4},10^{-3},0.005,0.01,0.0125$, with $\alpha=0$ and compression strains of $40\%$ and $60\%$. The high-alignment assays use $(v_0,\alpha)=(0.005,0.3)$ and target strains of $0\%$, $30\%$, and $60\%$. This representative point displays coherent rotation in the compressed solid and no comparable drift in the fluid; it is not a systematic map of the rotation boundary in the $(v_0,\alpha)$ plane.


In the supplied implementation, the ellipsoid center and semiaxes used for motility steering follow the cell-center extents until $t_{\rm load}+40$, after which they are fixed. This freeze time is prescribed relative to loading onset rather than determined by detecting the target strain.

The archived solid reference used for the twist analysis has $s_0=4.8$, $T_{eff}=10^{-3}$, $\Delta t=0.01$, $t_{\rm load}=540$, $t_{\rm end}=6500$, plate speed $u_{\rm p}=0.40$, and output interval 20. It retains $D_r=10^{-5}$, reconnection threshold $0.02$, $(v_0,\alpha)=(0.005,0.3)$, and a unit steering relaxation factor. The controlled release probes modify the noise and reconnection settings as described below.

The high-alignment studies consist of five runs per tissue and strain, with $s_0=4.8$ for the solid and $s_0=5.8$ for the fluid and $T_{eff}=10^{-3}$ in all 30 runs, including the compressed solid. Loading starts at $t=540$ with plate speed $0.40$; the steering ellipsoid is frozen at $t=580$ in the supplied implementation. The time-dependent rotation, shape, and strain-comparison panels normalize $t=579$--$5219$ to $[0,1]$, whereas Fig.~\ref{fig:stress_shape_alignment_dynamics}a--c uses the early $t=579$--$999$ window.

To convert simulation units to physical units, we assume a volume-equivalent cell diameter of $15\,\mu\mathrm m$ and a mean uncompressed three-dimensional speed of $2\,\mu\mathrm m\,\mathrm h^{-1}$. Matching these values to the high-activity fluid reference gives approximately $21.5$ hours per thousand model time units and $10.8$ hours for the 500-unit post-loading period of the zero-alignment series. These are assumed calibration values rather than an experimental fit, and the same model-time conversion is used across conditions. The simulations keep cell number fixed and omit cell division; the illustrative mapping spans hours for the short windows and several days for the long high-alignment trajectories.

The reference is the mean of eight pre-compression fluid runs at $v_0=0.0125$ assigned to $40\%$ strain: their three-dimensional centroid speed is $0.00349334$ and their volume-equivalent cell diameter is $1.216852$ in model units. These values give $0.021531$ hours per model-time unit under the stated assumptions. The high-alignment early window of 420 units corresponds to approximately $9.0$ hours, the 4640-unit normalized-time window to $99.9$ hours, and the full saved window from $t=579$ to $6499$ to $127.5$ hours on the same illustrative scale.

\subsection*{Cell motion, contact persistence, and shape observables}

A cell center is the arithmetic mean of its distinct vertex coordinates, with periodic-image corrections where required by the geometry representation. Cell velocity is obtained from center displacements between consecutive saved frames divided by their time separation. The in-plane speed is the magnitude of the $xy$ component, first averaged over cells and then over the specified sampling window. In the supplied zero-alignment analysis, the pre-loading speed window is $1000\le t<1400$, and the post-loading window begins at $t=1600$ and extends to the last available frame, capped at $t=2000$.

Of the 189 zero-alignment runs used in these figures, 186 have a final saved frame at $t=1999$; three fluid runs end at $t=1659$, $1819$, and $1899$. Fig.~\ref{fig:zero_alignment_compression_response}d uses endpoint $q_2$ values rather than time-window means; the speed changes retain the time-window averages defined above.

\subsubsection*{Contact persistence}
Both alignment datasets use the same contact-persistence criterion $q_2$. For each cell, the current neighbor set is compared with a fixed reference set: the cell contributes one if every reference neighbor is still present and zero otherwise; newly gained neighbors are ignored. The reported $q_2$ is the average of this binary indicator over the cells in the indicated population or radial layer. The high-alignment series was recomputed from saved cell--face topology at each sampled frame, using the same radial bins as the shape measurements. This indicator does not count individual T1 events or impose irreversible loss of persistence if an original neighbor is subsequently regained.

The zero-alignment code requests reference resets at $t=1000$ and $1500$; with the saved-frame schedule these occur at $t=1019$ and $1519$. The last pre-compression snapshot at $t=1499$ is compared with the $t=1019$ reference, and each run's final post-compression snapshot with the $t=1519$ reference. These endpoint values are used in Figs.~\ref{fig:zero_alignment_compression_response}d, \ref{fig:speed_neighbor_loss}, and \ref{fig:activity_neighbor_retention}. The high-alignment analysis instead uses the neighbor set at $t=579$ as its reference.

\subsubsection*{Cell shape and orientation}

For the high-alignment aspect ratio in Fig.~\ref{fig:cell_elongation_neighbor_stability}a, cell vertices are projected onto the $xy$ plane, their convex hull is constructed, and the covariance matrix of the hull-vertex coordinates is diagonalized. The aspect ratio is $\sqrt{\lambda_{\max}/\lambda_{\min}}$; the corresponding major eigenvector defines the projected long axis. This projected measure is distinct from the three-dimensional ratio $l_1/l_3$ used in Fig.~\ref{fig:zero_alignment_compression_response}c.

For the zero-alignment three-dimensional shape, the covariance of each cell's distinct vertex coordinates gives eigenvalues $\lambda_1\ge\lambda_2\ge\lambda_3$ and axis lengths $l_k=\sqrt{\lambda_k}$. Fig.~\ref{fig:zero_alignment_compression_response}c uses $l_1/l_3$, while Fig.~\ref{fig:zero_alignment_velocity_shape_correlation} uses the $xy$ projections of the corresponding major eigenvector and cell velocity. The long axes in Fig.~\ref{fig:stress_shape_alignment_dynamics}b--c instead come from a three-dimensional shape tensor of cell vertices and face centers; its major eigenvector is projected into $xy$.

For a cell at projected radius $r$ in the $xy$ plane, the local tangent is $\hat{\mathbf t}=(-y/r,x/r,0)$ and the radial direction is $\hat{\mathbf r}=(x/r,y/r,0)$. The tangential angle in Fig.~\ref{fig:stress_shape_alignment_dynamics}b is the angle between the projection of the cell's longest shape axis onto the $xy$ plane and $\hat{\mathbf t}$. Shape--velocity association is measured by $|\cos\psi_{xy}|$ between the $xy$ projections of the longest cell axis and center displacement. Its absolute value does not encode the sign of rotation.

Fig.~\ref{fig:stress_shape_alignment_dynamics}c uses the absolute cosine between the projected shape axis at one saved frame and the projected centroid displacement to the next saved frame, normally 20 model-time units later. Valid cell--frame values are pooled within each run and radial bin; the plotted points are means and standard errors over runs. The isotropic two-dimensional reference is $2/\pi$.

Signed angular velocities are calculated from wrapped changes of the cell-center polar angle between saved frames. The layer angular velocity is formed from cell-count-weighted radial-bin averages and integrated to obtain angular displacement. To combine runs with opposite rotation signs, each angular-displacement series is multiplied by the sign of its final displacement before averaging. Radial layers in the high-alignment analysis are defined by instantaneous projected radius: Inner, $r<1.5$; Mid, $1.5\le r<3$; and Outer, $r\ge3$.

\subsection*{Cell stress and radial averages}

The high-alignment analysis reads a scalar stress assigned to each cell and groups cells by their instantaneous projected radius. For Fig.~\ref{fig:radial_stress_early_si}, the supplied plotting script selects $579\le t\le1000$ and forms cell-count-weighted layer averages within each run. These run-level values are then averaged across five runs per tissue and strain.

The plotted maximum shear is $(\lambda_{\max}-\lambda_{\min})/2$ for the cell tensor $\boldsymbol\sigma_j=-2K_V(V_j-V_0)\mathbf I+[2K_A(A_j-A_0)/V_j](\mathbf M_j-A_j\mathbf I)$, where $\mathbf M_j=\sum_{f,e}\mathbf a_{fe}\mathbf a_{fe}^{\mathsf T}/|\mathbf a_{fe}|$ and $\mathbf a_{fe}$ is the triangular area vector formed by an edge and its face center. Thus the tensor is normalized by cell volume and includes volume and area elasticity, but no explicit boundary-tension, active, or plate-force term. Fig.~\ref{fig:stress_shape_alignment_dynamics}a and Fig.~\ref{fig:radial_stress_early_si} select the same saved frames, $t=579$--$999$.

\subsection*{Statistics and uncertainty estimation}
All response quantities are per-run means, averaged over cells and over the sampling window of the run; error bars on binned or grouped data are standard errors of the mean over runs. The endpoint-$q_2$ and speed association uses 189 runs across 24 tissue--activity--strain conditions; one fluid trajectory lacks valid final geometry for geometry-dependent analyses. Because replicates share both protocol and initial conditions, we quote \emph{cluster-robust} standard errors with the simulation condition as the cluster, rather than conventional least-squares errors. Unless stated otherwise, $\pm$ denotes one cluster-robust standard error and $p$-values are two-sided.

\subsection*{Local, five-cell energy-barrier calculations}

The ordered and disordered local, five-cell geometries provide mechanical reference calculations for the susceptibility to cellular rearrangements.  They are separate from the active full-spheroid dynamics. 

The ordered constructions use regular truncated octahedra as its reference geometry. Each initially undeformed cell has six square faces, eight hexagonal faces, and 24 vertices. Kim, {\it et al.}\cite{kim2024mean} analyzed this space-filling geometry using a vertex-model energy with quadratic penalties for deviations from target volume and surface area. The undeformed regular truncated octahedron has a geometric shape index $s_{\rm TO}=A/V^{2/3}\simeq5.31474$. To estimate the energy barrier associated with a cell-neighbor rearrangement, we construct a five-cell cluster from a body-centered-cubic Voronoi tessellation. The cluster consists of the three cells sharing a selected edge and the two additional cells associated with its endpoints. The endpoints of a selected edge are displaced symmetrically toward one another and its length is progressively reduced for each strain step. 

At each prescribed edge length, the remaining vertices are relaxed collectively by minimizing the total vertex-model energy. We set \(V_0=1\) and \(K_V=1\), and vary the preferred surface area \(A_0\), which is equivalent to the target shape index \(s_0\) for unit-volume cells. The minimization is performed using the L-BFGS-B algorithm with analytically calculated energy gradients. The relaxed configuration from the preceding step is used as the initial configuration for the next minimization, implementing a quasistatic edge-shrinking protocol. The optimizer uses a relative function tolerance of \(10^{-10}\), a gradient tolerance of \(10^{-6}\), and a maximum of 5000 iterations per minimization.

The exterior boundary of the five-cell cluster is left unconstrained, allowing neighboring cells to deform collectively as the selected edge shrinks. The rearrangement barrier is determined from the maximum increase in relaxed energy relative to the initial configuration along this prescribed edge-shrinking pathway. A weak harmonic tether is introduced to eliminate a residual rigid-body rotational degree of freedom. Although the endpoints of the shrinking edge are constrained, the cluster can still rotate about the axis connecting these endpoints without changing its internal surface-area or volume energy. To suppress this rotation, one additional vertex away from the shrinking edge is elastically tethered to its reference position with coupling constant, $k_{tet}$. The tether serves only as a numerical regularization rather than a physical constraint on the cells. We use a small tether stiffness, $k_{tet}=10^{-2}$ and verify that the calculated rearrangement barrier is insensitive to the tether strength for several orders of magnitude.

The following protocol applies to the disordered five-cell calculations. Each disordered five-cell realization is extracted from a periodic Voronoi tessellation generated from 1000 independently and uniformly distributed seeds in a cubic box of side length 10. Candidate control edges have both endpoints within a distance of 0.5 from the box center; the longest candidate is selected. The cells incident on its endpoints form the five-cell cluster, with three cells sharing the edge and one additional cell at each endpoint. Twenty stored topologies are reused across the parameter sweep. Each combination of target shape index and imposed strain starts by reloading its topology, rather than continuing from a preceding strain or edge scan. The resulting statistics concern this selected-edge ensemble.

The cluster carries the same vertex-model energy as the spheroid, $E=\sum_j(A_j-A_0)^2+K_V\sum_j(V_j-V_0)^2$ with $K_V=1$, no surface tension, activity, or alignment, and $V_0$ and $A_0$ (or equivalently $s_0=A_0/V_0^{2/3}$) set to the cluster's initial mean volume and surface area.

Before relaxation, the cluster is rotated so that the plane through the area-weighted surface centers of the three edge-sharing cells becomes parallel to the $xy$ plane. The two loading series then compress along $z$ or $x$, respectively normal or tangent to this reference plane. This construction does not impose an exact control-edge angle after relaxation. Initial unconstrained relaxation, post-compression relaxation, and every scan-point relaxation use L-BFGS-B with analytic forces and three starts: the current coordinates and two Gaussian perturbations of standard deviation 0.03 in the free coordinates. The lowest-energy result is retained. The optimizer uses a projected-gradient tolerance of $10^{-5}$, a relative energy-change tolerance of $10^{-8}$, and a maximum of $10^7$ iterations. These are optimizer stopping settings, rather than an independently enforced residual-force acceptance test.

Pre-compression is performed by moving two virtual plates from $\pm H_0$ to $\pm H_0(1-\varepsilon)$, with $H_0$ set by the relaxed cluster's extent along the loading axis. The supplied production drivers use 200 nominal loading steps and an explicit overdamped time step of 0.005. A vertex outside a plate experiences an inward normal force with stiffness 80 and force cap 20, in addition to its internal force; each vertex displacement is capped. A detected outward crossing of a plate triggers rollback of the coordinate update and reduction of the plate increment. No intermediate energy minimization is requested during loading. After loading, every vertex recorded as contacting a plate is placed at its final plate position and fixed in the normal direction, while its tangential coordinates remain free; the other vertices obey hard bounds within the final plate gap. The cluster is then minimized under these constraints. Thus the plate-contact set is inherited from the loading history. These vertices cannot detach normally during the post-loading relaxation; during scanning, the same constraints remain on the non-control vertices, while the two control endpoints follow their prescribed positions.

The control edge is scanned from its post-compression length $L_{\rm init}$ at fixed midpoint and direction. Its two endpoints are fixed at symmetrically displaced positions for each minimization, while the remaining coordinates are relaxed under the post-compression constraints. The production drivers use 30 sampled lengths, including the initial length; the final sample has length $2\times10^{-6}L_{\rm init}$. Connectivity is held fixed throughout preparation, compression, and scanning: neither the control edge nor background edges undergo a reconnection. The calculation therefore measures a prescribed edge-collapse branch and does not implement the subsequent edge-to-face transition.

For each topology, the reported barrier is $\Delta E=\max_k E_k-E_0$, where $E_0$ is the minimized energy of the first scan point at the original post-compression edge length. A reported zero means that no sampled energy exceeds $E_0$; there is no separate rule assigning zero when an edge collapses during loading. Subprocess failures are recorded as missing values, whereas an unsuccessful minimizer termination produces a warning and does not by itself exclude the result. The supplied aggregate tables contain twenty finite barriers at each of the 77 combinations of $s_0=4.8,4.9,\ldots,5.8$ and $\varepsilon=0,0.1,\ldots,0.6$ for each loading axis. The phase diagram applies the $10^{-4}$ threshold to the topology-averaged barrier. This threshold equals the effective noise energy scale used in the zero-alignment full-spheroid simulations but does not establish an active rearrangement rate.

The five-cell calculation uses $K_V=1$, whereas the full spheroid uses $K_V=10$. Its fixed connectivity, prescribed scan coordinate, and loading-history constraints define a simplified mechanical comparison rather than a parameter-matched measurement of the active spheroid.

The twenty input topologies and individual energy paths for all 1540 topology--parameter combinations per loading direction are retained; their barriers underlie the reported medians and fractions below $10^{-4}$. The production output does not retain per-path force-residual, face-intersection, or relaxed control-edge-angle diagnostics. Accordingly, $z$ and $x$ identify loading normal and tangential to the aligned middle-cell plane, not exact orientations relative to the control edge.

\subsection*{Differential-twist probes and heuristics}

Static twist stiffness is evaluated on five matched solid configurations at saved time $t=799$ for each of $\epsilon=0,0.3,0.6$. This reference assay uses $s_0=4.8$, $K_V=10$, $K_A=1$, and $\Gamma=1$. Coordinates are centered on the mean position of vertices belonging to real cells. With $\rho_i$ their cylindrical radius and $R_{xy}$ the 99th percentile of these vertex radii, we construct an azimuthal displacement proportional to $(-y_i,x_i,0)(\rho_i/R_{xy})^2$. Its equal-vertex-weight projection onto rigid rotation and its mean translation are removed. Displacements outside the real-vertex set are zero, and the real-vertex root-mean-square displacement is normalized to unity. The resulting mode is denoted by $\mathbf d$.

With topology fixed, the stiffness is the projected central force difference
\[
K_{\rm twist}=-\frac{\mathbf d\cdot[\mathbf F(\mathbf x+h\mathbf d)-\mathbf F(\mathbf x-h\mathbf d)]}{2h\,\mathbf d\cdot\mathbf d},
\]
where $\mathbf F$ is the full reconstructed mechanical force and $h=5\times10^{-5}$. Coordinates outside the prescribed displacement remain fixed, and no additional coordinate minimization is performed. Comparison with $h=10^{-4}$ gives a maximum relative difference below $6\times10^{-7}$ for the fifteen snapshots used here. Figure~\ref{fig:twist_response_high_alignment}a reports the full stiffness, without subtracting the uncompressed baseline.

Release probes use controlled restarts derived from the saved $t=800$ geometry of one matched topology at each strain. The original saved polarity fields are unavailable, so the same isotropic random polarity initialization is assigned across strains. Paired perturbations are applied to the subsequent active restart state at restart time 99. Each state is displaced by $\pm a\mathbf d$, with $a=0.02$, using identical initial polarities within each pair and across its active/passive comparison. Active probes use $v_0=0.005$ and passive probes $v_0=0$, both retaining $\alpha=0.3$. Translational and rotational noise are set to zero, and reconnections are suppressed by setting their length threshold to zero. The integration step is $0.01$. For the compressed probes, the plates return to the archived target gap at speed $0.40$ and are then held fixed; the zero-strain control has no additional plate motion. These controlled preparations are not exact continuations of the original active trajectories.

The normalized twist response is obtained from the paired trajectories as
\[
q(t)=\frac{[\mathbf x^+(t)-\mathbf x^-(t)]\cdot\mathbf d}{2a\,\mathbf d\cdot\mathbf d},
\]
where the projection is restricted to the real-vertex mode, giving $q(0)=1$. Probe time starts at release. The rates used in the heuristic estimate are fitted to $\ln q(t)=c-\lambda t$ over $0.5\le t\le5$ with a free intercept, using the 10-unit probes. Figure~\ref{fig:twist_response_high_alignment}b shows the $60\%$ probes extended to 50 model time units and plots $t/50$; saved points extend to $t=49$. These longer responses are not single exponentials, and the early-time rates are not fits over the entire displayed curve.

The mean stiffnesses at zero and $60\%$ strain are $0.29818$ and $0.25619$. A linear approximation over the measured range is
\[
K_{\rm tw}(\epsilon)\simeq0.298-0.070\epsilon.
\]
The zero-strain passive fit gives $\lambda_{\rm tw}(0)=0.085699$, while the $60\%$ passive and active fits give $0.066243$ and $0.062391$. Their difference is $0.003852$. The heuristic calibration $\mu_{\rm tw}=\lambda_{\rm tw}(0)/K_{\rm tw}(0)\simeq0.287$ combines a five-geometry stiffness mean with a single controlled-restart decay rate. Approximating this effective mobility as strain independent gives
\[
\lambda_{\rm tw}(\epsilon)\simeq0.0857-0.0201\epsilon,
\qquad
\epsilon_c\simeq\frac{0.0857-\Lambda_{\rm act}}{0.0201}.
\]
The constant-mobility approximation is used for this estimate and is not imposed on the measured release curves. At $60\%$ strain, it gives $\lambda_{\rm tw}(0.6)\simeq0.0736$, compared with the directly fitted passive rate of $0.066243$. The calibration therefore gives an approximate feedback requirement: growth of the proposed coupled mode at this strain would require $\Lambda_{\rm act}$ of order $0.074$. The measured active--passive rate difference quantifies the effect of activity on the controlled relaxation; it does not measure the full feedback combination $c_vc_\alpha\alpha v_0/\lambda_p$. Determining the rotation threshold requires that feedback measurement and resolution of onset across strains.

For the loading-stress estimate, plate pressure is reconstructed as the mean of the top and bottom reaction-force magnitudes divided by $\pi R_{\rm c,99}^2$, where $R_{\rm c,99}$ is the 99th percentile of real-cell-center cylindrical radii. This cell-center radius differs from the vertex radius used to normalize the twist mode. At $t=799$, the secant modulus is the mean matched difference $E_{\rm sec}(\epsilon)=\langle[P_i(\epsilon)-P_i(0)]/\epsilon\rangle$ across the five topologies. It gives $E_{\rm sec}(0.6)=1.56587$ and a plate-pressure increment of $0.93952$, or, in rounded form,
\[
\Delta\sigma_{\rm plate}(0.6)\simeq E_{\rm sec}(0.6)\times0.6\simeq0.94
\]
in simulation stress units, with $E_{\rm sec}(0.6)\simeq1.57$. This matched-pressure measurement establishes the axial loading scale independently of the constant-mobility approximation. It is separate from the cell-level scalar stress used in the radial profiles and is evaluated at the selected strain rather than at a measured rotation threshold.

\begin{acknowledgments}
Tao Zhang acknowledges financial support from the NSFC/China via award 22303051. The computations in this paper were run on the $\pi$ 2.0 and the Siyuan-1 cluster supported by the Center for High Performance Computing at Shanghai Jiao Tong University. JMS acknowledges financial support from the National Science Foundation under Grant No.~PoLS-2412961.
\end{acknowledgments}

\bibliography{ref}
\clearpage
\onecolumngrid

\begin{center}
{\large\bfseries Supplementary Information\par}
\vspace{0.6em}
{\Large\bfseries Active tissue mechanics drives a diversity of collective flows in uniaxially compressed 3D spheroids\par}
\vspace{0.8em}
{\normalsize Liyang Wang,$^{1}$ Shabeeb Ameen,$^{2}$ J. M. Schwarz,$^{2,3}$ and Tao Zhang$^{1,*}$\par}
\vspace{0.5em}
{\small\itshape
$^{1}$School of Chemistry and Chemical Engineering, Shanghai Jiao Tong University, Shanghai 200240, China\par
$^{2}$Physics Department, Syracuse University, Syracuse, NY 13244, USA\par
$^{3}$Indian Creek Farm, Ithaca, NY 14850, USA\par}
\vspace{0.4em}
{\small $^{*}$zhangtao.scholar@sjtu.edu.cn\par}
\end{center}
\vspace{0.8em}

\renewcommand{\thefigure}{S\arabic{figure}}
\setcounter{figure}{0} 

\begin{figure}[!htbp]
\centering
\includegraphics[width=0.5\textwidth]{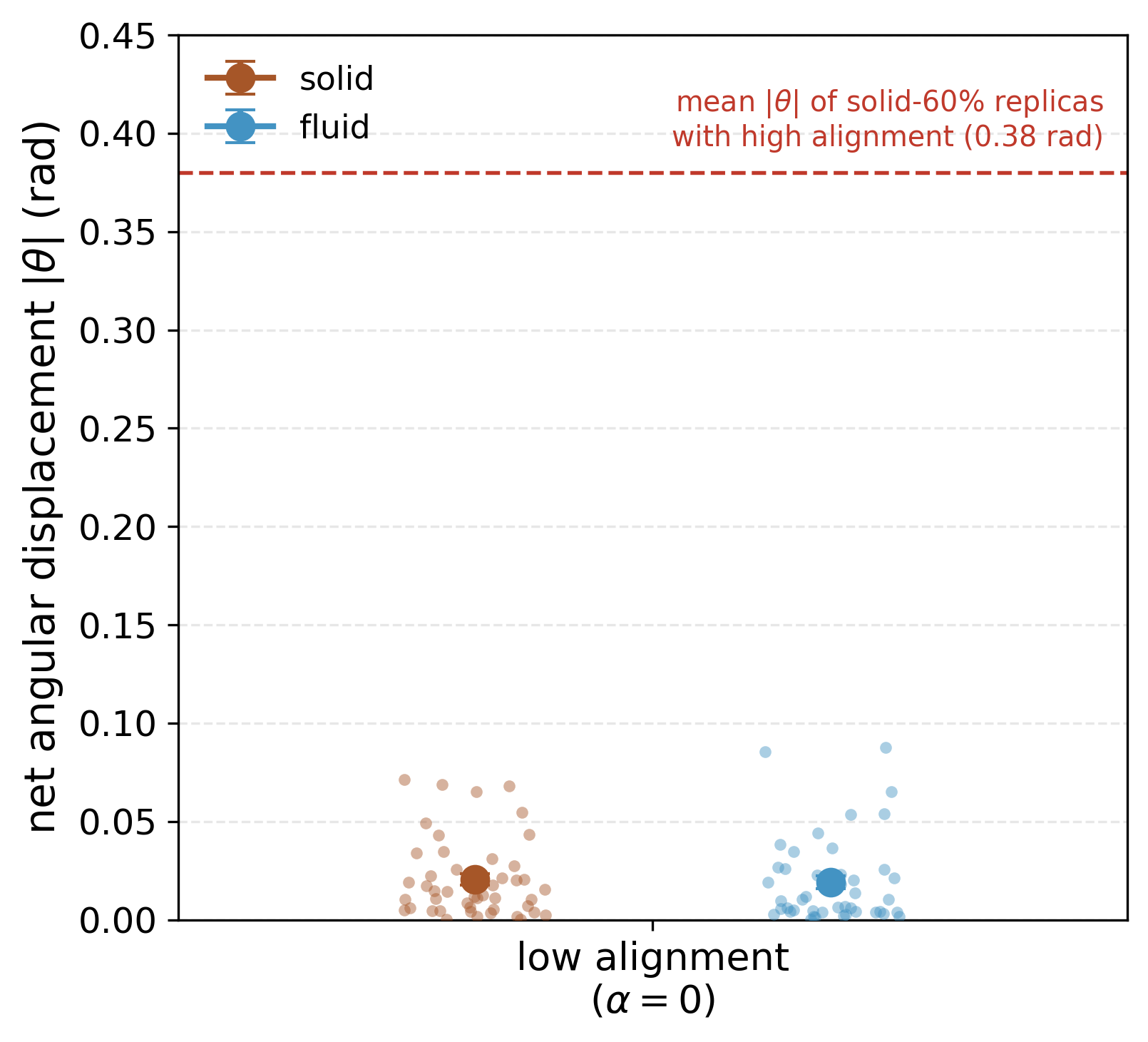}
\caption{{\it No coherent rotation at low alignment.}  Net angular displacement
magnitude $|\theta|$ accumulated after the $60\%$ compression pulse, per
low-alignment run ($\alpha=0$) of the Section~\ref{sec:lowalign} sweeps, for
solid ($s_0=5.0$) and fluid ($s_0=5.6$) spheroids (small points: individual
runs; large markers: mean $\pm$ SEM).  Every low-alignment run remains below
the mean net displacement of the five solid-60\% replicates measured with
high alignment ($0.38$~rad, dashed line). Post-pulse tracking extends up to $480$ model time units, whereas the high-alignment reference spans $4640$ units. The two series also use different parameter sets (Methods); this figure compares accumulated displacements over their respective observation windows.}
\label{fig:zero_alignment_angular_displacement}
\end{figure}

\begin{figure}[!htbp]
\centering
\includegraphics[width=\textwidth]{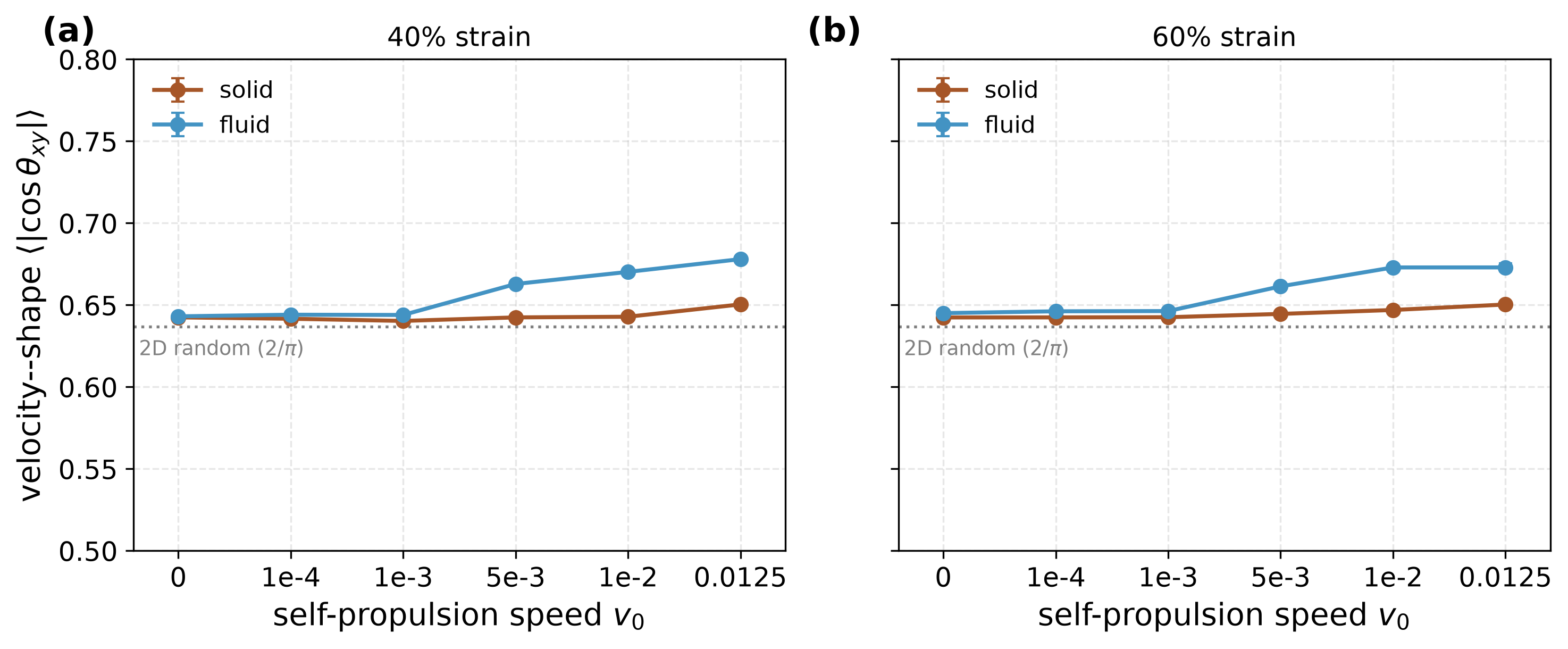}
\caption{{\it Shape--velocity correlation with zero alignment.}
Mean velocity--shape correlation $\langle|\cos\theta_{xy}|\rangle$ between each
cell's projected $xy$ velocity and projected longest shape axis, averaged over all cells and
time, versus active speed $v_0$, for solid
($s_0=5.0$, brown) and fluid ($s_0=5.6$, blue) spheroids at (a) $40\%$ and
(b) $60\%$ strain (mean $\pm$ SEM over replicates; the highest-activity fluid at $60\%$ strain has five runs and the other groups have eight; same runs as
Fig.~\ref{fig:zero_alignment_compression_response}). Dotted line: isotropic two-dimensional random expectation $2/\pi$. At zero alignment the correlation remains near this reference across the sweep and rises modestly in the most active fluids. Together with the small net angular displacements (Fig.~\ref{fig:zero_alignment_angular_displacement}), the near-isotropic shape--velocity association characterizes the disordered motion in this series.}
\label{fig:zero_alignment_velocity_shape_correlation}
\end{figure}

\begin{figure}[!htbp]
\centering
\includegraphics[width=\textwidth]{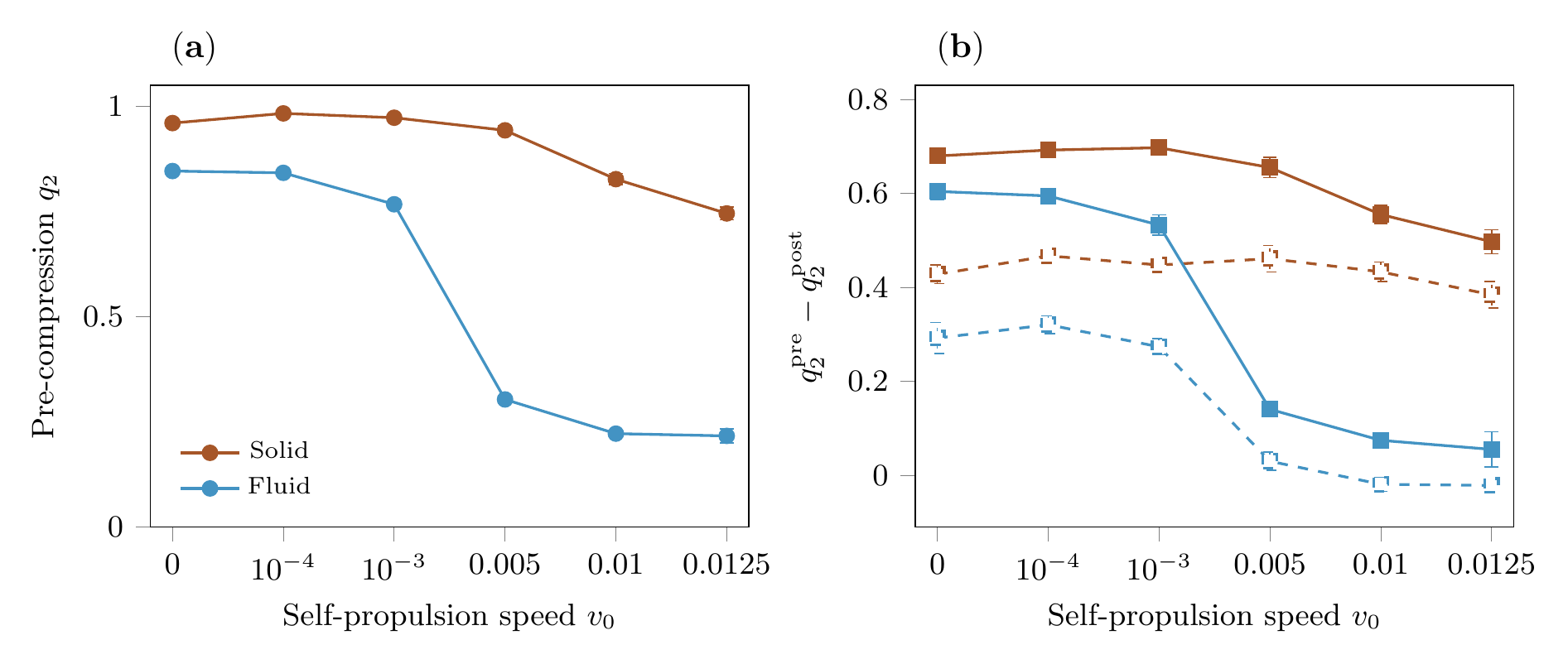}
\caption{{\it Endpoint neighbor retention across activity and strain.} (a) Pre-compression $q_2$ versus active speed $v_0$ (mean $\pm$ SEM over replicates, strains pooled). Contact retention decreases strongly with activity in the fluid, while the solid retains more neighbors throughout the sweep. (b) Endpoint neighbor-retention loss, $q_2^{\rm pre}-q_2^{\rm post}$, versus activity, shown separately at $40\%$ and $60\%$ strain. At $v_0=0.0125$, the fluid means are $-0.022$ and $0.055$, respectively. These endpoint differences measure retention of neighbor sets rather than the number of rearrangement events.}
\label{fig:activity_neighbor_retention}
\end{figure}

\begin{figure}[!htbp]
\centering
\includegraphics[width=\textwidth]{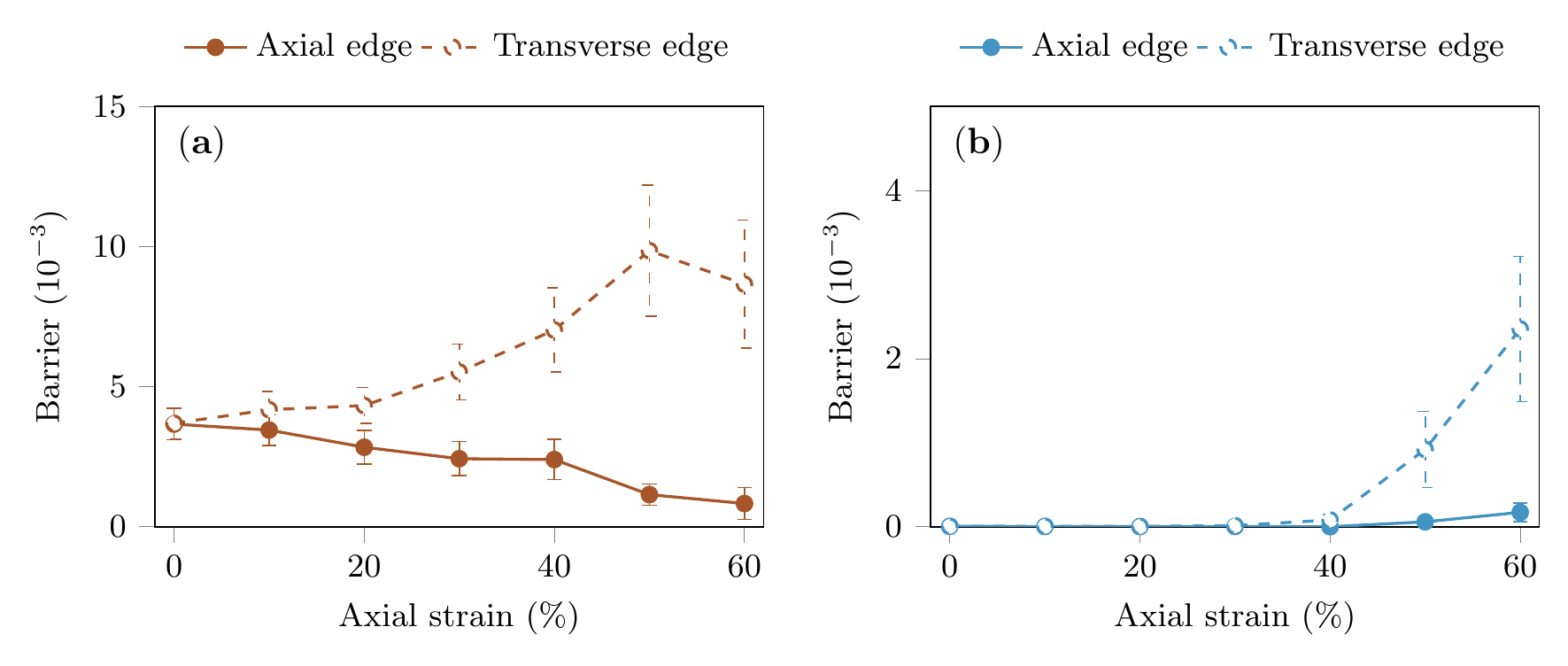}
\caption{{\it Five-cell barrier calculations.} A prescribed control-edge shrinkage under $z$- and $x$-directed loading in the aligned cluster frame, respectively normal to and within the initial middle-cell plane. These protocols do not constrain the final control-edge angle. Panels (a) and (b) show $s_0=5.0$ and $s_0=5.6$, respectively; points are means $\pm$ SEM over twenty topologies, including zero-barrier paths. The high-strain response depends on loading direction: the fluid mean under $x$ loading rises while its median remains zero at $60\%$, showing that the increased mean coexists with barrierless paths.}
\label{fig:directional_barriers_si}
\end{figure}

\begin{figure}[!htbp]
\centering
\includegraphics[width=\textwidth]{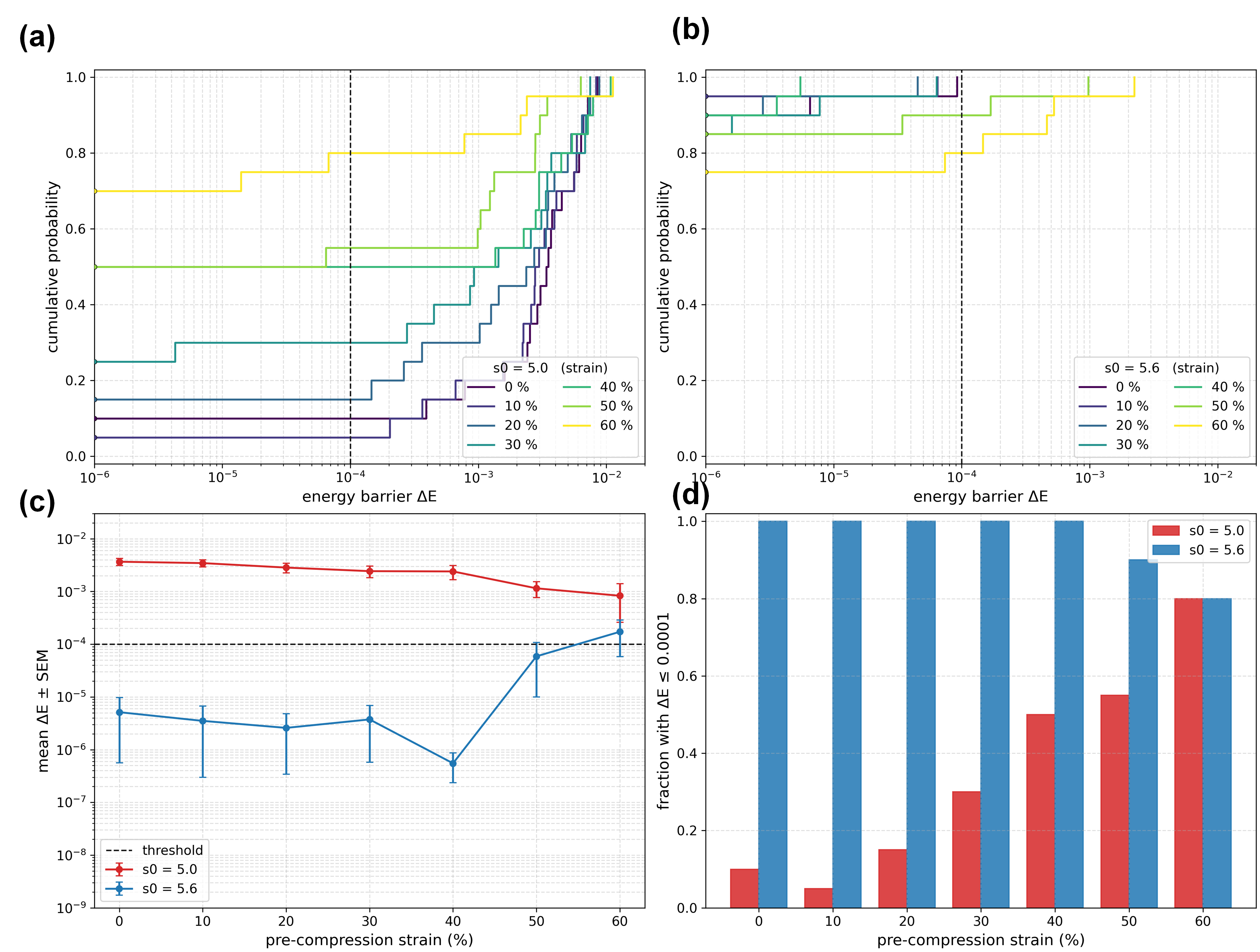}
\caption{Distribution of the prescribed edge-collapse energy barrier $\Delta E$ under $z$-direction pre-compression for $s_0=5.0$ (brown) and $s_0=5.6$ (blue), with twenty independent realizations per condition. (a,b) Empirical cumulative distributions for strains from $0\%$ to $60\%$ at $s_0=5.0$ (a) and $s_0=5.6$ (b); the dashed line marks $\Delta E=10^{-4}$. (c) Mean barrier ($\pm$ SEM) versus strain. The $s_0=5.0$ mean decreases from $3.7\times10^{-3}$ to $8.3\times10^{-4}$, whereas the $s_0=5.6$ mean falls from $30\%$ to $40\%$ strain and rises at higher strain to $1.7\times10^{-4}$ at $60\%$. The high-strain fluid increase is driven by the high-barrier tail, with a zero median at $60\%$. (d) Fraction of realizations with $\Delta E\le10^{-4}$. For $s_0=5.0$ this fraction increases overall, with a small initial decrease; for $s_0=5.6$ it remains near unity through $40\%$ strain and decreases at higher strain. Both fractions reach $0.8$ at $60\%$, so low-barrier paths remain common even when the mean exceeds the threshold. Zero-barrier realizations are retained in all statistics and appear as a single jump at the left edge of the logarithmic axis.}
\label{fig:t1_barrier_distribution}
\end{figure}

\begin{figure}[!htbp]
\centering
\includegraphics[width=\textwidth]{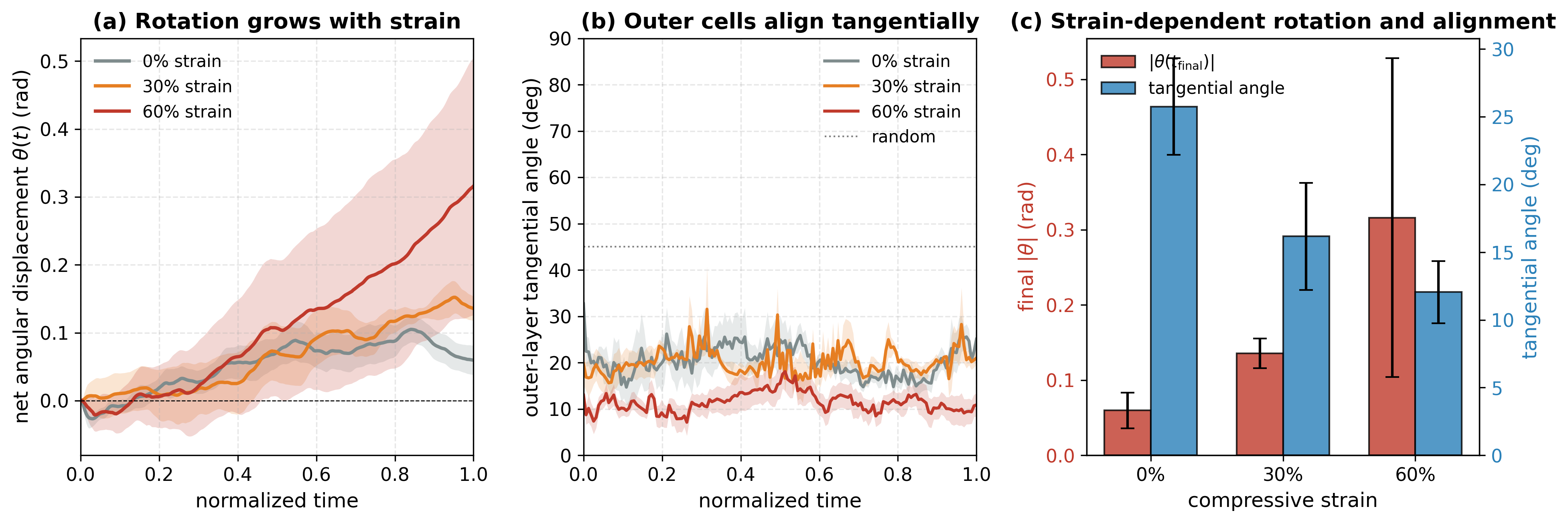}
\caption{Compression-dependent rotation and peripheral cell orientation in solid spheroids.  (a) Net angular displacement $\theta(t)$ (chirality-normalized,
mean $\pm$ SEM over replicas) for solid spheroids at $0\%$, $30\%$ and
$60\%$ strain, integrated over the same window used in Fig.~\ref{fig:collective_rotation_dynamics}
.  (b) Outer-layer tangential angle of cell long axes versus time
(mean $\pm$ SEM over replicas; computed from the pre-averaged mean-correlation
field in the radiusMetrics output as $90^\circ - \mathrm{correlation}$ and taking the
outermost radius bin per frame); the $45^\circ$ line marks a random orientation.
(c) Mean final rotation magnitude increases and the corresponding mean outer-layer tangential angle decreases across the three tested strains. These endpoint trends link stronger compression to greater accumulated rotation and stronger peripheral tangential organization. Results use $n=5$ replicas per condition.}
\label{fig:strain_dependent_rotation_alignment}
\end{figure}

\begin{figure}[!htbp]
\centering
\includegraphics[width=0.75\textwidth]{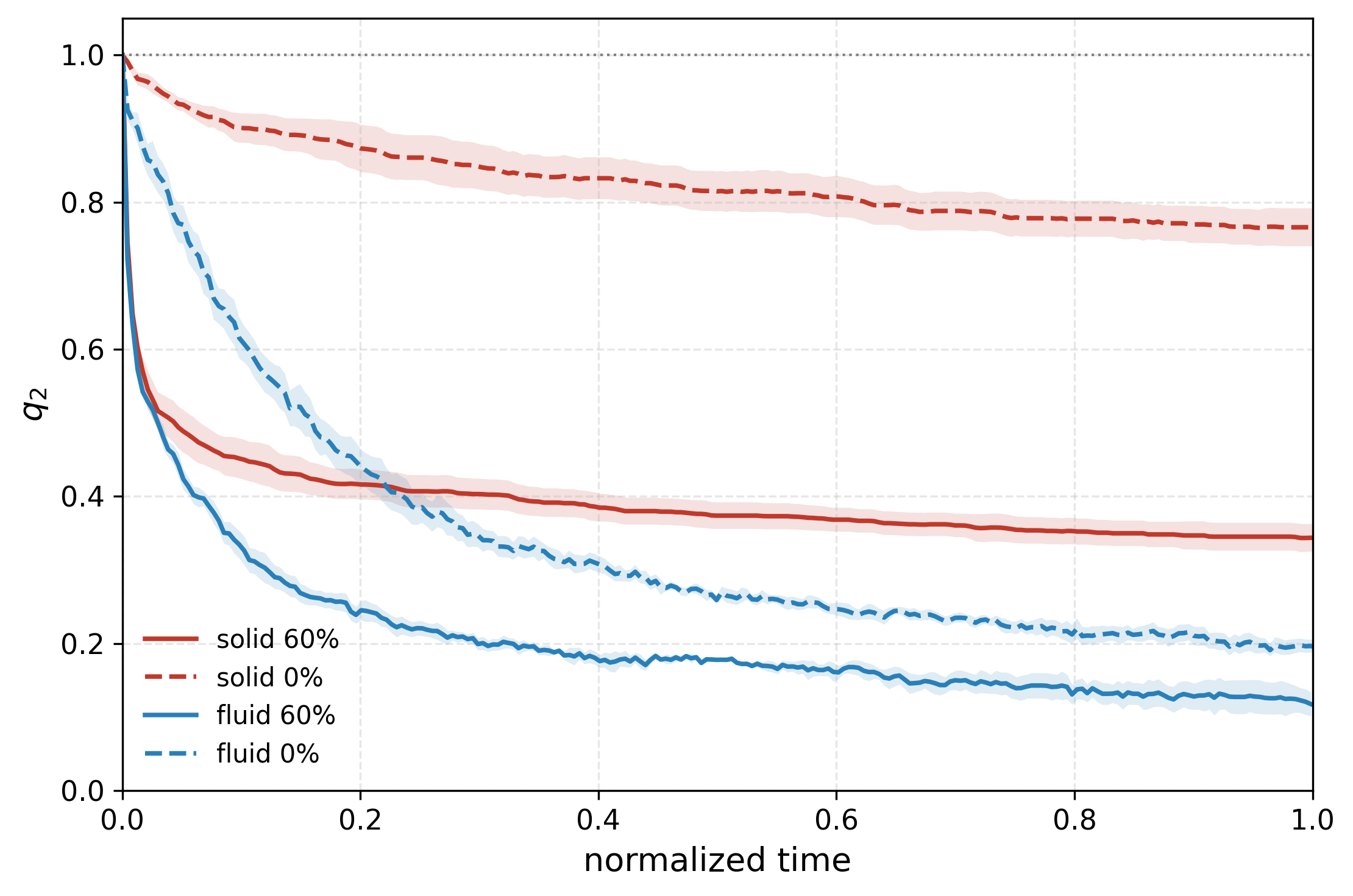}
\caption{Contact-persistence index $q_2$ versus normalized time for fluid-like and solid-like spheroids at $0\%$ and $60\%$ strain (mean $\pm$ SEM over replicates). A cell contributes one if every neighbor in its reference set at $t=579$ is still present; newly gained neighbors are ignored (Methods). At $0\%$ strain the solid maintains relatively high contact persistence, with a final value of approximately $0.77$. At $60\%$ strain its index drops initially after loading and then decays more slowly, remaining above the fluid's.}
\label{fig:neighbor_retention_relaxation}
\end{figure}

\begin{figure}[!htbp]
\centering
\includegraphics[width=\textwidth]{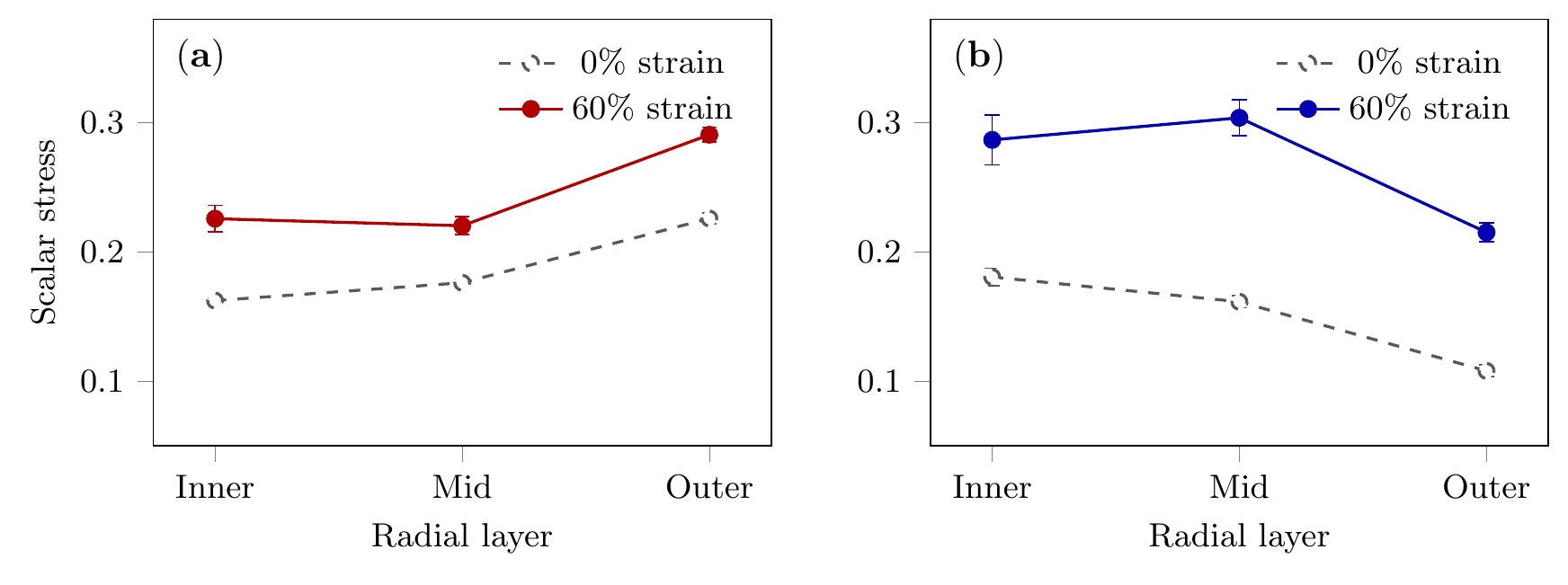}
\caption{Radial profile of the scalar cell-stress measure during the early window $579\le t\le1000$ at $0\%$ and $60\%$ target strain. Panels (a) and (b) show solid and fluid spheroids, respectively. Each point is a mean of five run-level, cell-count-weighted layer averages; error bars are SEM across runs. At $60\%$ strain, the solid outer layer has the highest scalar stress, whereas the fluid outer layer lies below its inner and middle layers. The plotted quantity is the scalar stress measure defined in Methods.}
\label{fig:radial_stress_early_si}
\end{figure}

\end{document}